# Ultrabroad resonance of localized plasmon on a nanoparticle coupled with surface plasmon on a nanowire enabling two-photon excited emission via continuous-wave laser

Tamitake Itoh[1*], Yuko S. Yamamoto[2]

[1]Health and Medical Research Institute, National Institute of Advanced Industrial Science and Technology (AIST), Takamatsu, Kagawa 761-0395, Japan

[2]School of Materials Science, Japan Advanced Institute of Science and Technology (JAIST), Nomi, Ishikawa 923-1292, Japan

*Corresponding author: tamitake-itou@aist.go.jp

**ABSTRACT**

This study found that plasmonic hotspots (HSs) between silver nanoparticles (NPs) and silver nanowires (NWs) generated two-photon excited emissions, including hyper-Rayleigh, hyper-Raman, and two-photon fluorescence of dye molecules with continuous-wave (CW) near-infrared (NIR) laser excitation. A comparison between experimental results and electromagnetic (EM) calculations revealed that a large EM enhancement factor ($F_R$) at the HS appears in the visible to NIR regions owing to EM coupling between localized plasmons of the NP and surface plasmons of the NW. This ultrabroad resonance of the coupled plasmons enables the observation of two-photon-excited emissions with CW laser excitation. A comparison of the dependence of the NP diameter, NW diameter, and gap distance on two-photon-excited emissions further demonstrated that the large HS-by-HS variations in the emission intensities were mainly induced by the sensitivity of $F_R$ to the gap distance at the HSs. We also experimentally and theoretically investigated the propagation of two-photon-excited emission light to a neighboring NP on the NW via surface plasmons.

# I. Introduction

Nonlinear optical spectroscopy is a powerful tool for investigating molecular electronic and vibronic properties that are not observable by linear spectroscopy because of the complementary selection rules of photo-induced transitions.[1] The excited polarization $P_{in}$ of a molecule is expressed as a power series in the amplitudes of the incident electric field amplitude $E_{in}$ as:

$$P_{in} = \varepsilon_0 \left[ \chi^{(1)} E_{in} + \chi^{(2)} E_{in}^2 + \chi^{(3)} E_{in}^3 + \cdots \right] = P_{in}^{(1)} + P_{in}^{(2)} + P_{in}^{(3)} + \cdots, \quad (1)$$

where $\chi^{(2)}$ and $\chi^{(3)}$ with $P_{in}^{(2)}$ and $P_{in}^{(3)}$ are the second- and third-order nonlinear susceptibilities with excited polarizations, respectively.[1] By assuming $\chi^{(1)}$ as unity, the typical values of $\chi^{(1)}$, $\chi^{(2)}$, and $\chi^{(3)}$ are as small, as 1, 2 × 10$^{-12}$ m/V, and 4 × 10$^{-24}$ m$^2$/V$^2$.[1] Thus, nonlinear spectroscopy usually requires ultrafast laser pulses with high peak $E_{in}$ values to detect $P_{in}^{(n)}$ to compensate for the extremely small nonlinear susceptibilities.[1] Such strong field excitation sometimes induces the decomposition of target molecules via multi-photon absorption or the photothermal effect.[2] Furthermore, the use of ultrafast laser pulse light, the line-width of which is much wider than that of continuous wave (CW) laser light,[3] may deteriorate the spectral resolution.[4] Thus, nonlinear spectroscopy with weak $E_{in}$ without using an ultrafast laser pulse is expected to resolve these issues. The electromagnetic (EM) enhancement via the local plasmon (LP) resonance of metallic nanoparticles (NPs) is useful to resolve the issue.[4-6] In the case of the spontaneous de-excitation rate of a molecule interacting with the NP, its enhancement is expressed by the Purcell effect as $F = \dfrac{Q(\lambda/n)^3}{4\pi^2 V_P}$, where $Q$ is the quality factor of the LP resonance; $\lambda$ and $n$ are the light wavelength and refractive index around the NP, respectively; and $V_P$ is the mode volume of the LP resonance.[7] The EM enhancement of excitation or emission rates $F_R$ is expressed as the radiative portion of $F$:

$$F_R = \dfrac{F \Delta \omega_R}{\Delta \omega_R + \Delta \omega_{NR}} \left( = \left| \dfrac{E_{loc}}{E_{in}} \right|^2 \right), \quad (2)$$

where $\Delta \omega_R$ and $\Delta \omega_{NR}$ are the radiative and nonradiative decay rates of the LP resonance, respectively; and $E_{loc}$ is the amplitude of the enhanced local electric field.[1] The maximum $F_R$ reaches approximately $10^5$ at the nanogaps or junctions between metallic NP dimers owing to the extremely small $V_P$.[7-9] Such places are referred to as plasmonic hotspots (HSs). HSs have received considerable attention because they exhibit various interesting phenomena, including single molecule surface-enhanced Raman scattering (SERS),[10-13] ultrafast surface-enhanced fluorescence (ultrafast SEF),[14] and strong coupling between plasmons and molecular excitons.[15,16] Various plasmon modes, including radiant, non-radiant, and sub-radiant plasmons have been also investigated to optimize the plasmon-enhanced spectroscopies at HSs.[17,18] In the case of surface enhanced nonlinear spectroscopy, the $n$th-order nonlinear excitation polarization $|P^{(n)}|^2$ in Eq. (1) exhibits an enhancement factor of $F_R^n < (10^5)^n$.[1] Thus, the total enhancement, including the emission enhancement of the $n$th-order nonlinear photoemission $F_{SEnPE}$, becomes:

$$F_{SEnPE}(\lambda_{ex}, \lambda_{em}) = F_R(\lambda_{ex})^n F_R(\lambda_{em}), \quad (3)$$

where $\lambda_{ex}$ and $\lambda_{em}$ are the excitation and emission wavelengths for harmonic generation, respectively.[5,6] The maximum value of $F_{SEnPE}$ in Eq. 2, that is, approximately $(10^5)^{n+1}$,

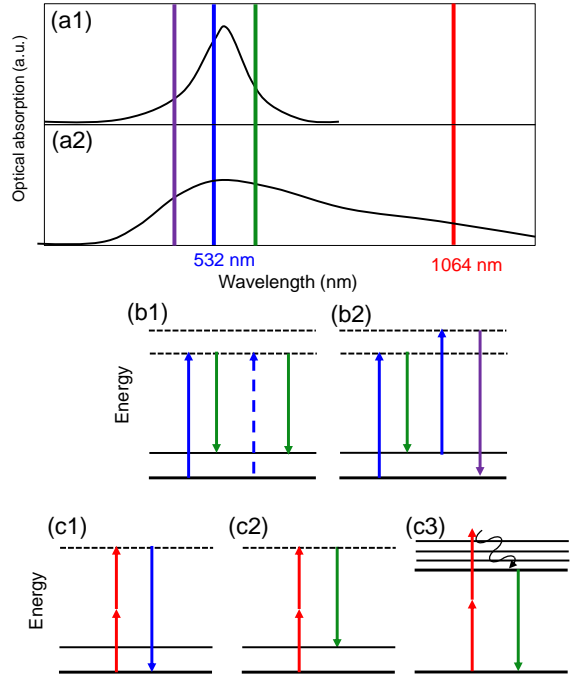

FIG. 1. (a1) Optical absorption spectrum of the LP resonance of NP. (a2) Absorption spectrum of coupled resonance between the LP of NP and SP. Purple, blue, green, and red lines correspond to coherent anti-Stokes Raman, hyper-Rayleigh (or visible laser line), stimulated Raman, and NIR laser lines. (b1) and (b2) Energy-level diagrams for photo-induced transitions of stimulated Raman scattering and coherent anti-Stoke Raman scattering, respectively. (c1)–(c3) Energy-level diagrams of hyper-Rayleigh scattering, Stokes hyper-Raman scattering, and two-photon fluorescence, respectively.

implies nonlinear spectroscopy, even with CW laser excitation at the HSs.[19,20]

The spectral linewidths of the LP resonance of NPs or their dimers are limited to less than one or two hundred nanometers, as shown in Fig. 1(a1).[21] Thus, in the case of $\lambda_{ex} \sim \lambda_{em}$, such as that observed in surface-enhanced stimulated Raman scattering and surface-enhanced coherent anti-Stokes Raman scattering,[4,22] the value of their EM enhancement factors becomes large, as both the $\lambda_{ex}$ and $\lambda_{em}$ lines are within the linewidth of the LP resonance, as shown in Figs. 1(a1), 1(b1), and 1(b2). However, for $\lambda_{ex} \gg \lambda_{em}$, such as in the surface-enhanced hyper Rayleigh scattering (SEH-Ray), surface-enhanced hyper Raman scattering (SEHRS), and two-photon surface enhanced fluorescence (two-photon SEF) shown in Figs. 1(c1)–1(c3),[5,6,19,20] the generation of $F_{SE2PE}$ becomes inefficient because the $\lambda_{ex}$ and $\lambda_{em}$ lines cannot be within the common LP resonance linewidth as shown in Fig. 1(a1). Thus, we consider that the coupled resonance between the LP and surface plasmon (SP) is useful for obtaining a broad resonance in which the linewidth includes both $\lambda_{ex}$ and $\lambda_{em}$, as shown in Fig. 1(a2), because the energy of the SP continues from the ultraviolet region to zero.[23,24] The coupled resonance between the LP of the NP and SP of the NW extends the resonance linewidth from the visible to the near-infrared (NIR) region.[23]

In this study, we investigated the generation of surface-enhanced two-photon emission (SE2PE), such as SEH-Ray, SEHRS, and two-photon SEF, of dye molecules at the HSs between silver NPs and silver NWs with CW-NIR laser excitation. We also measured surface-enhanced one-photon emission (SE1PE), such as SERS and SEF, at common HSs. The HSs exhibited several interesting optical properties. The Rayleigh scattering spectra of the NPs composing the HSs exhibited vibrational structures that did not appear in the spectra of the isolated NPs. The SE1PE and SE2PE intensities exhibited considerable HS-by-HS variations and uncorrelated intensities. The polarization extinction ratios of the SE2PE were always larger than those of the SE1PE. The propagation of SE2PE light to neighboring NP through the NW was frequently observed. The EM calculations reproduced these optical properties well, indicating that the origin of these phenomena is the large $F_R$ appearing from the visible to NIR regions by coupling between the LPs of the NPs and the SPs of the NWs.

**II. Experiment**

The silver NW colloidal dispersion was prepared as described previously.[25] The silver NPs were the by-products inside the NW colloidal dispersion. Figures 2(a1) and 2(a2) show the scanning electron microscopy (SEM) images and NW diameter ($D_{NW}$) distribution, respectively. The average NW diameter and length were approximately 70 nm and approximately 10 μm, respectively. Figures 2(b1) and 2(b2) show the SEM image and the NP diameter ($D_{NP}$) distribution, respectively. The distribution of the NP diameters was approximately 100–240 nm. Moreover, NPs usually deviate from their spherical shapes.[26]

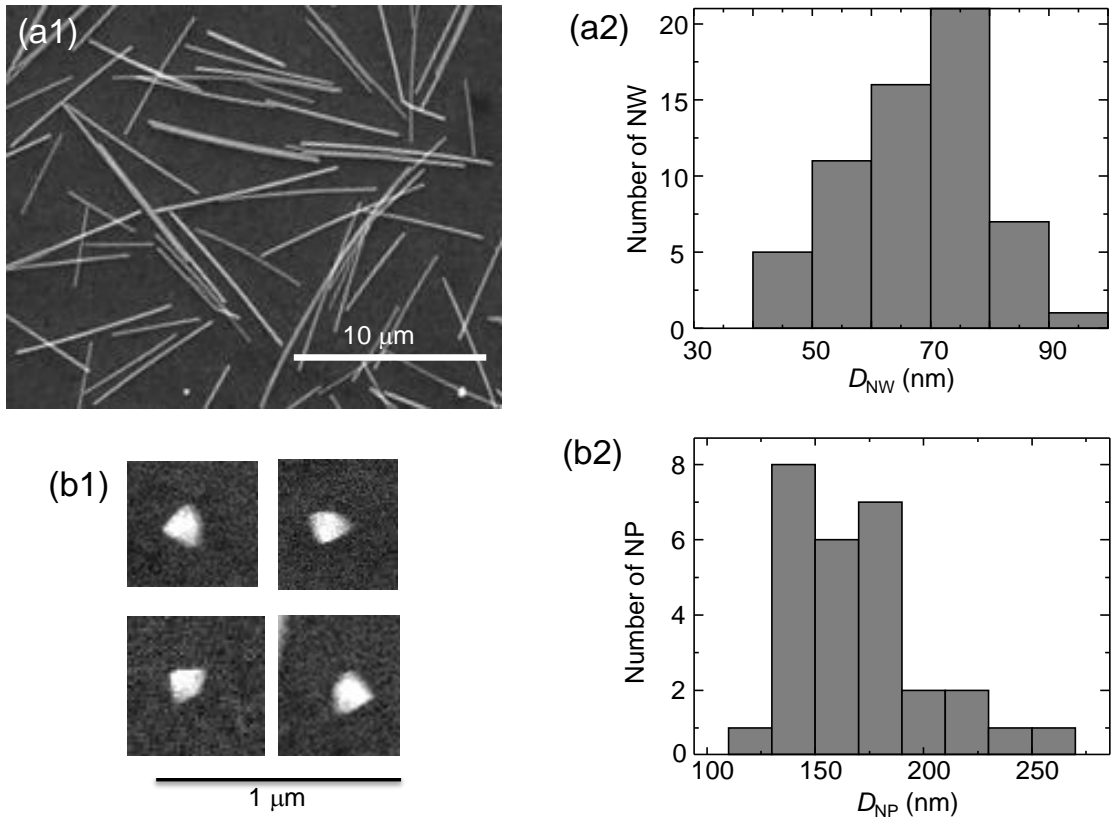

FIG. 2. (a1) SEM image of NWs. Scale bars represent 10 μm. (a2) Distribution of $D_{NW}$ estimated via SEM images. (b1) SEM images of NPs. Scale bars represent 1 μm. (b2) Distribution of $D_{NP}$ estimated by SEM images.

A mixture of the dispersion of NWs, including NPs and rhodamine 6G (R6G) methanol solution (approximately $5.0 \times 10^{-6}$ M) was dropped and dried on a glass plate. Thereafter, the effective concentration of the dye on the NWs was reduced by photobleaching most of the excess dye molecules adsorbed on the NWs, NPs, and glass surfaces via green laser beam excitation for approximately 10 min. Photobleaching was confirmed by the disappearance of fluorescence from the glass surface.[27]

The Rayleigh scattering of isolated NWs and NPs was examined using the finite-difference time-domain (FDTD) method. Details of the FDTD calculations are described in Sec. III. Figures 3(a1) and 3(a2) show the experimental and calculated Rayleigh scattering spectra of the NW edges. The $D_{NW}$ in the calculation was set to 70 nm, based on Fig. 2(a2). The vertical axis of the experimental Rayleigh scattering spectrum is expressed as a ratio between the Rayleigh scattering intensity of single NW edges or NPs and that of large NP aggregates corresponding to the white light source spectrum.[18] This vertical axis represents Rayleigh scattering efficiency.[18] The vibrational structures in both experimental and calculated spectra are results of the Fabry–Pérot (FP) interference of the SP propagating through the NW, indicating that the NWs can be treated as single crystalline rather than polycrystalline.[28] Figures 3(b1) and 3(b2) show the experimental and calculated Rayleigh scattering spectra of single silver NPs. The $D_{NP}$ in the calculation was set to 50–200 nm, based on Fig. 2(b2). Most of the NPs exhibited broad peaks around <600 nm in the experiments shown in Fig. 3(b1), indicating that the effective value of $D_{NP}$ is approximately 200 nm, based on the calculated spectra shown in Fig. 3(b2).

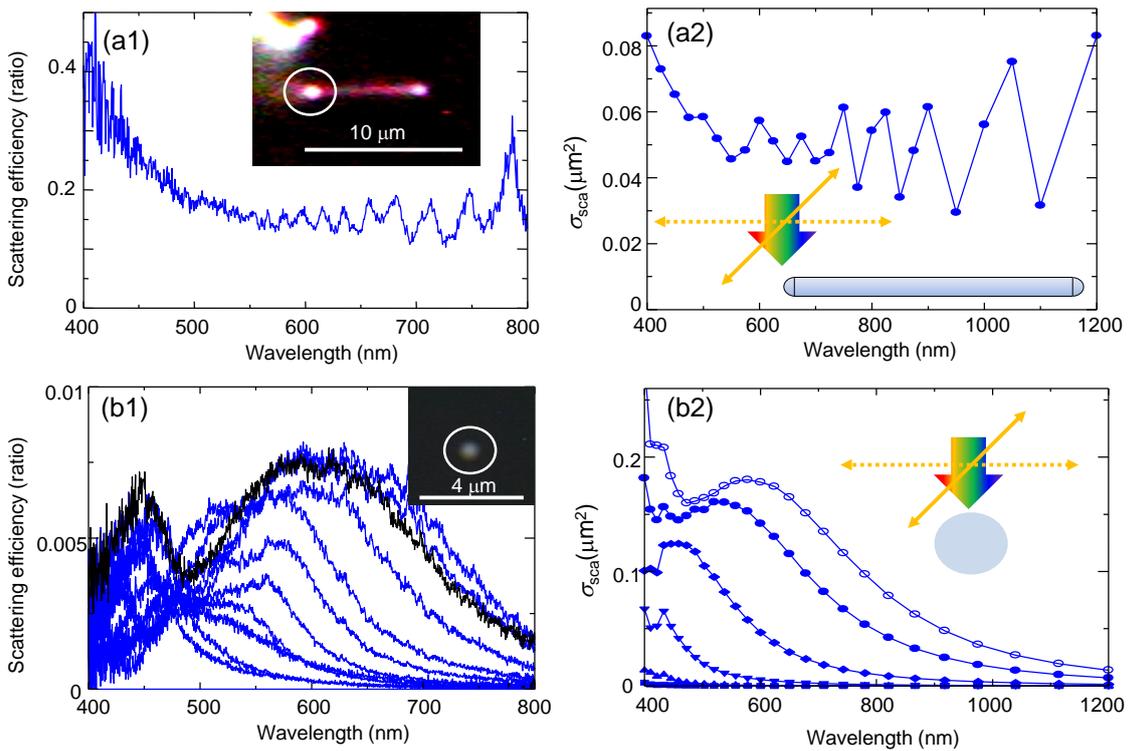

FIG. 3. (a1) Experimental Rayleigh scattering spectrum from the edge of the isolated NW, as indicated by the circle in the dark-field image. Scale bars represent 10 μm. (a2) Calculated Rayleigh scattering spectrum from the edge of a single NW with a $D_{NW}$ of 70 nm and length of 6 μm, respectively, excited at the edge of the illustration. (b1) Experimental Rayleigh scattering spectra of isolated 12 single NPs indicated via a dark-field image of the black spectrum. Scale bars represent 4 μm. (b2) Calculated Rayleigh scattering spectra of single NPs with a $D_{NP}$ of 50–200 nm, excited as illustrated.

The experimental setups for detecting the spectra of Rayleigh scattering, SE1PE (including SERS and SEF), and SE2PE (including SEHRS, SHE-Ray, and two-photon SEF) are shown elsewhere.[20] The Rayleigh scattering spectrum of a NP on a NW placed on a glass plate was measured via illumination through a dark-field condenser [numerical aperture (NA) 0.92] using a 50 W halogen lamp as a white light source. The NPs on the NWs could be identified by their dark and vivid colored spots in the bright- and dark-field images, respectively, as shown in Figs. 4(a1) and 4(a2). The SE1PE spectra of the HSs between the NPs and NWs were measured under wide-field excitation (approximately $200 \times 300$ μm$^2$) using a green laser beam (DPSS laser, 532 nm). The laser beam was focused onto the sample using a lens (NA = 0.2). The power density of the laser beam was 35 W/cm$^2$ at the focal point. Figure 4(b1) shows a SE1PE image of the HS. The NP regions shown in Figs. 4(a1) and 4(a2) exhibit SE1PE activity, as shown in Fig. 4(b1). The SE2PE spectra of the HSs were measured with the narrow-field excitation (approximately $0.65 \times 0.65$ μm$^2$) of a CW-NIR beam (cw-Nd3+:YAG laser, 1064 nm).

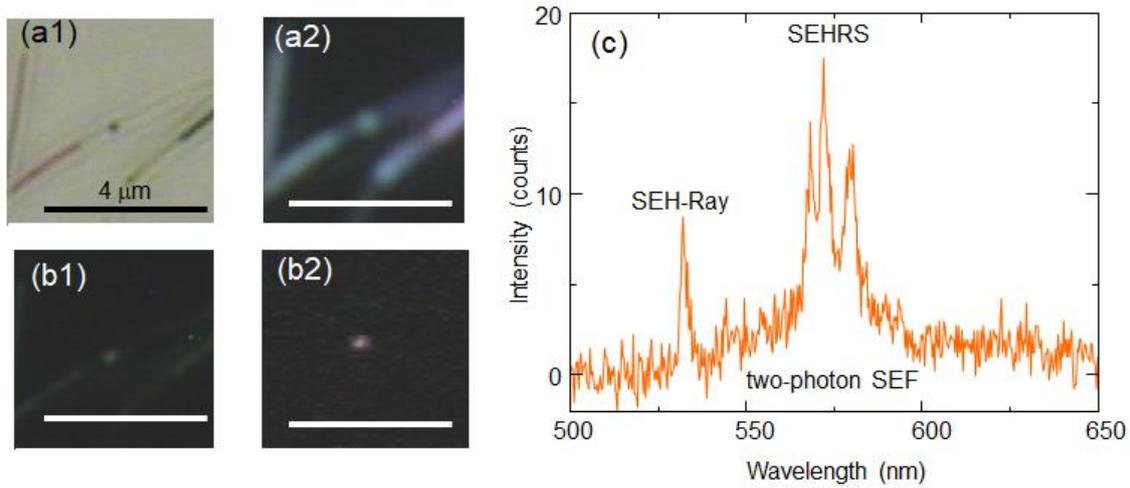

FIG. 4. (a1) and (a2) Bright- and dark-field images of a HS composed of NP and NW, respectively. (b1) and (b2) SE1PE and SE2PE images of the HS of NP on NW, respectively. All scale bars represent 4 μm. (c) Typical SE2PE spectrum composed of SEH-Ray and SEHRS lines with a two-photon SEF background.

The NIR laser beam was focused onto the NP on the NW using an objective lens (×100, NA 1.3). The HS at the NP in the NW region in Fig. 4(b1) exhibited SE2PE activity, as shown in Fig. 4(b2). Figure 4(c) shows a typical SE2PE spectrum, which is composed of the SEH-Ray line at 532 nm, SEHRS lines at approximately 550–580 nm, and a two-photon SEF as a broad background.

**III. Results and discussion**

Figure 5(a1)–5(a3) show SEM, dark-field, and SE1PE images, respectively. These images directly demonstrate the HS formed at the junction between the NP and NW and are consistent with the results of previous studies.[26,29-31] Figure 5(b) shows the typical

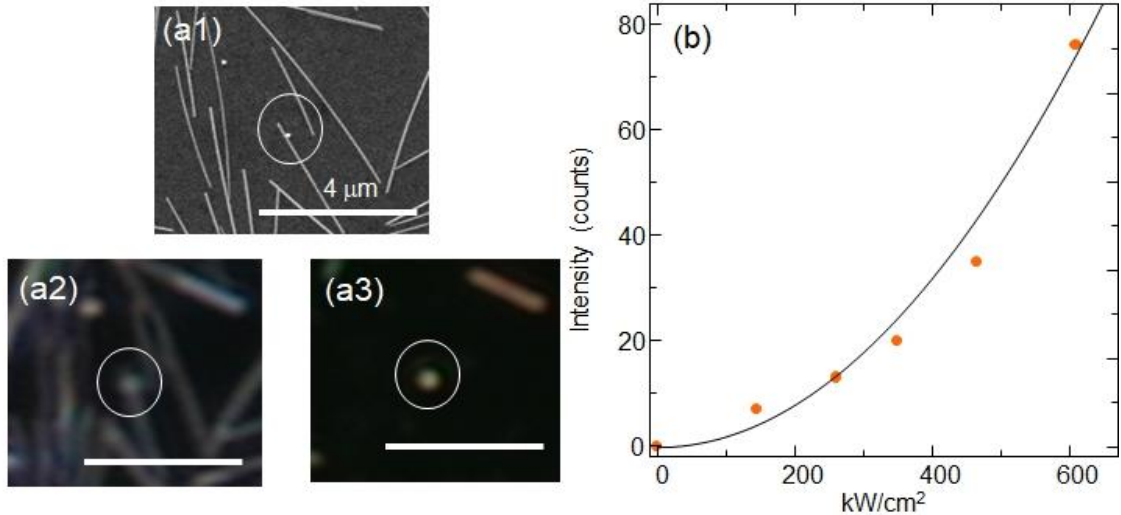

FIG. 5. (a1)–(a3) SEM, dark-field, and SE1PE images of HS composed of a NP and a NW marked by white open circles. All scale bars represent 4 μm. (b) Typical incident NIR laser power dependence of SE2PE intensity (ochre closed circles) with a fitted quadratic line (black solid curve) against the incident NIR laser power.

NIR laser power density dependence of the SE2PE light intensity fitted with a quadratic function. This fitting provided evidence that the observed spectra in Fig. 5(b) were induced by $P_{in}^{(2)}$, as shown in Eq. (1). This quadratic dependence also verifies the physical stability of the HS under the current laser power density because the exponent of the fitting function is decreased by the physical instability of the HS.

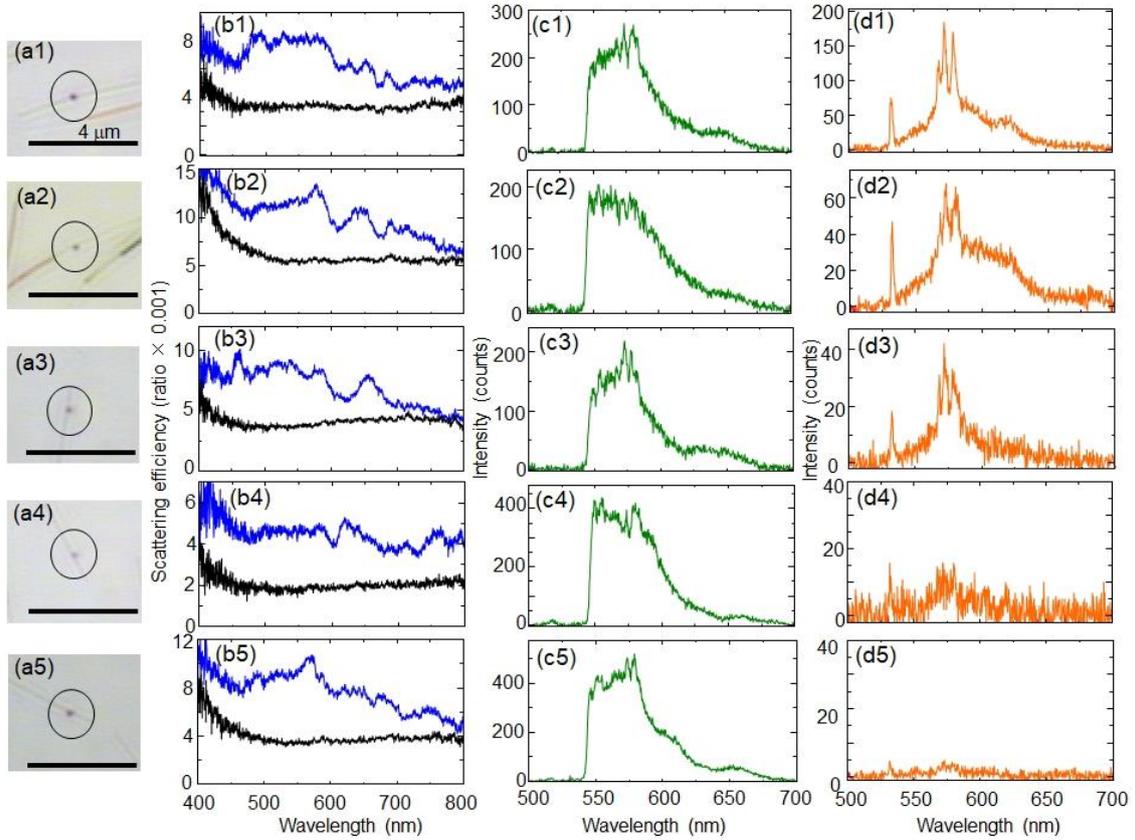

FIG. 6. (a1)–(a5) Bright-field images of HSs composed of NPs on NWs. The HSs are marked by black open circles. All scale bars 4 μm. (b1)–(b5) Rayleigh scattering spectra (blue curves) of the HSs and those (black curves) of NWs outside the HSs. (c1)–(c5) and (d1)–(d5) SE1PE (green curves) and SE2PE (ochre curves) spectra of the HSs.

Figures 6(a1)–6(a5), 6(b1)–6(b5), 6(c1)–6(c5), and 6(d1)–6(d5) show the typical data for the bright-field images of the HSs as well as the spectra of the Rayleigh scattering,

SE1PE, and SE2PE at the HSs, respectively. The NPs adsorb on NWs in Figs 6(a1)–6(a5). The Rayleigh scattering spectra of the HSs in Figs 6(b1)–6(b5) are largely different from the isolated NPs in Fig. 3(b1). These complex structures in their spectral shapes may be attributable to the effect of the FP interference of the SPs, as shown in Fig. 3(a1).[27,28] The spectral features of FP interference are transferred to a LP resonance of NP via EM coupling between NPs and NWs. These spectral features were examined via FDTD calculations. Figures 6(c1)–6(c5) show that the SE1PE spectra are mainly composed of SEF, thus indicating that R6G molecules close to HSs additionally contribute to SE1PE because the quenching effect in SEF outside HSs is much weaker than those inside HSs.[32] Figures 6(d1)–6(d5) show that the relative two-photon SEF intensities to SEHRS intensities appear smaller than the relative SEF intensities to SERS intensities. This tendency indicates that R6G molecules generating SE2PE receive more considerable quenching than do those generating SE1PE. This indication means that the molecules solely inside HSs contribute to SE2PE. The spectral variations in SE1PE and SE2PE have been explained as the spectral modulation by the second EM enhancement factor $F_R(\lambda_{em})$ in Eq. (2), for which spectral shapes have been derived from SEF spectra.[5,33] The complex structures in the Rayleigh scattering spectra in Figs. 6(a1)–6(a5) always appear for HSs showing SE1PE activity in Figs. 6(b1)–6(b5); however such HSs sometimes do not show

SE2PE activity, as shown in Fig. 6(d5). These results indicate that the generation of SE2PE is more tightly restricted than the EM coupling, which generates the complex structures in Rayleigh scattering spectra and the SE1PE.

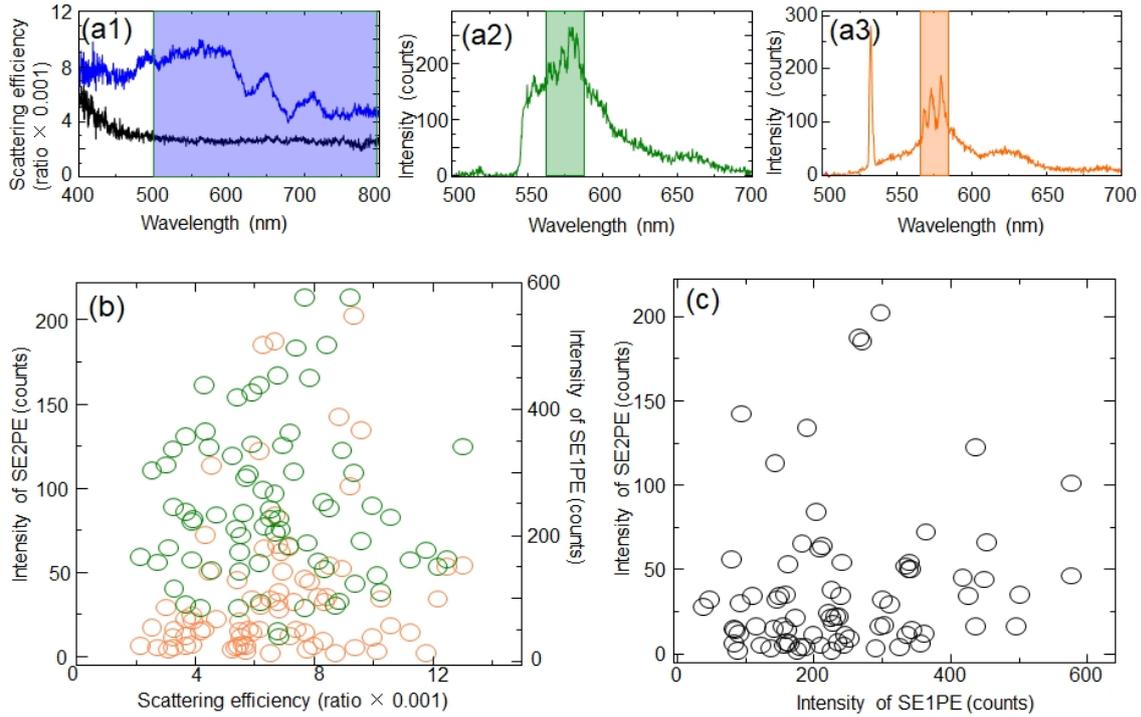

FIG. 7. (a1)–(a3) Rayleigh scattering (blue curve), SE1PE (green curve), and SE2PE (ochre curve) spectra of a HS composed of a NP on a NW. The average intensities of the covered areas are defined as Rayleigh scattering, SE1PE, and SE2PE intensities. (b) Relationship between Rayleigh scattering and SE1PE intensities (green open circles) of the 83 HSs and relationship between Rayleigh scattering and SE2PE intensities (ochre open circles) of the HSs. (c) The relationship between SE1PE and SE2PE intensities (black open circles) of the 83 HSs.

We explored the relationship between the Rayleigh scattering, SE1PE, and SE2PE intensities. These intensities were derived from the spectral areas indicated in Figs. 7(a1)–(a3). Figure 7(b) shows the SE1PE (and SE2PE) intensities versus the Rayleigh scattering intensities for the 83 HSs. No clear tendencies were observed for these relationships. The distribution of SE2PE intensities appears to be more compressed around the zero count than around the distribution of SE1PE intensities. These results are attributable to the nonlinearity in the first enhancement factors for SE2PE. That is, the first enhancement factors of SE2PE, $F_R(1064\text{ nm})^2$, spread the intensity distribution more than do that of SE1PE, $F_R(532\text{ nm})$. Figure 7(c) shows the SE2PE intensities versus SE1PE intensities for the 83 HSs. No clear tendencies were observed in these relationships. These results may be attributable to differences in the EM enhancement factors, as expressed by Eq. (2). Moreover, the EM enhancement factor of SE2PE is a product of $F_R(1064\text{ nm})^2$ and $F_R(\sim 550\text{ nm})$.[5] In contrast, the EM enhancement factor of SE1PE is a product of $F_R(532\text{ nm})$ and $F_R(\sim 550\text{ nm})$.[5]

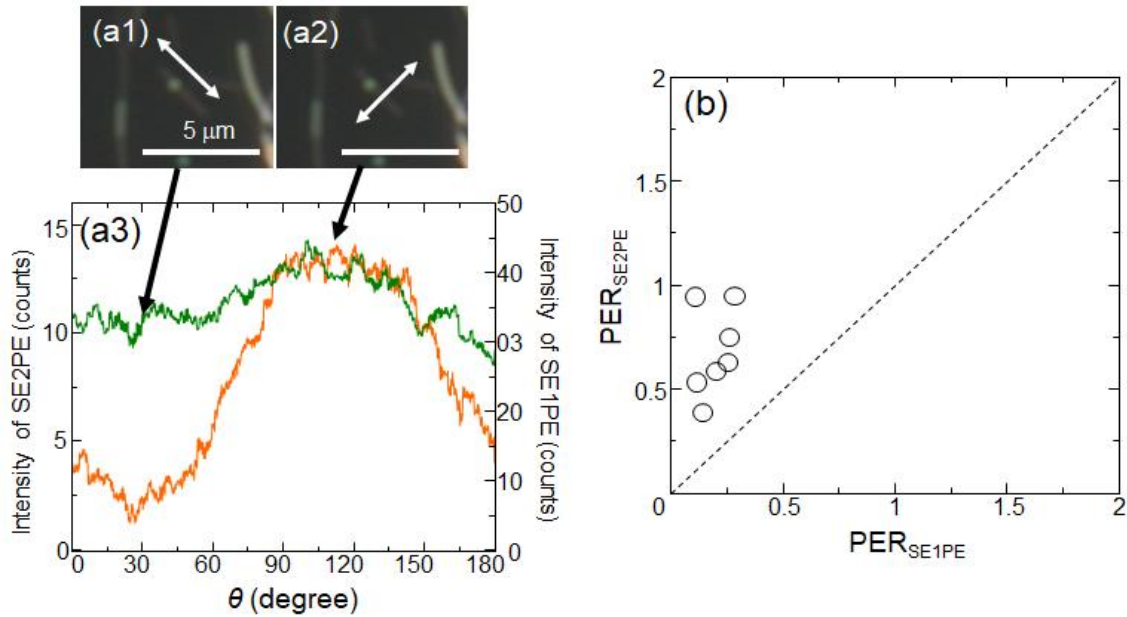

FIG. 8 (a1) and (a2) Dark-field images of a NP on a NW, showing $\theta$ for detection angles using white arrows. All scale bars are 5 μm. (a3) Polarization angle dependences of SE1PE (green curve) and SE2PE (ochre curve) intensities. (b) Relationship between $PER_{SE2PE}$ and $PER_{SE1PE}$.

We examined the polarization properties of the SE1PE and SE2PE intensities. Figures 8(a1) and 8(a2) show the polarization directions parallel and perpendicular to the long NW axis, respectively. Figures 8(a3) shows the polarization angle ($\theta$) dependence of SE1PE and SE2PE intensities. Both were measured by rotating the polarizer inserted in front of the detector. Both dependencies were fitted to $\cos^2\theta + A$ curves, where $A$ is a constant. The $\theta$ values exhibiting the maximum intensity for the SE1PE and SE2PE light are common and correspond to the perpendicular to the NW long axis. These $\theta$ dependences indicate that both SE1PE and SE2PE are mainly generated by the LP resonance, the direction of which is perpendicular to the long NW axis. The results for SE1PE are consistent with those of a previous report.[26,29,35] The minimum SE1PE intensity does not reach zero at a value of $\theta$ parallel to the NW long axis; even the SE2PE intensity approaches zero at $\theta$. This inconsistency was examined using polarization extinction ratios ($10 \times \log_{10}(I_{max}/I_{min})$), where $I_{max}$ and $I_{min}$ are the maximum and minimum intensities for SE1PE and SE1PE as $PER_{SE1PE}$ and $PER_{SE2PE}$. Figure 8(b) shows that $PER_{SE1PE}$ was always larger than $PER_{SE2PE}$ for all seven HSs. The second EM enhancement, $F_R$(~560 nm), was almost the same for SE1PE and SE2PE. Thus, the property $PER_{SE2PE} > PER_{SE1PE}$ may be induced by the difference between the polarization properties of $F_R$(1064 nm)$^2$ in SE2PE and $F_R$(532 nm) in SE1PE. Regarding the large average $D_{NP} \sim 200$ nm, as shown in Fig. 2, $F_R$(1064 nm) and $F_R$(532 nm) may be attributable to the dipole and quadrupole LPs, respectively,[18] resulting in different polarization properties.

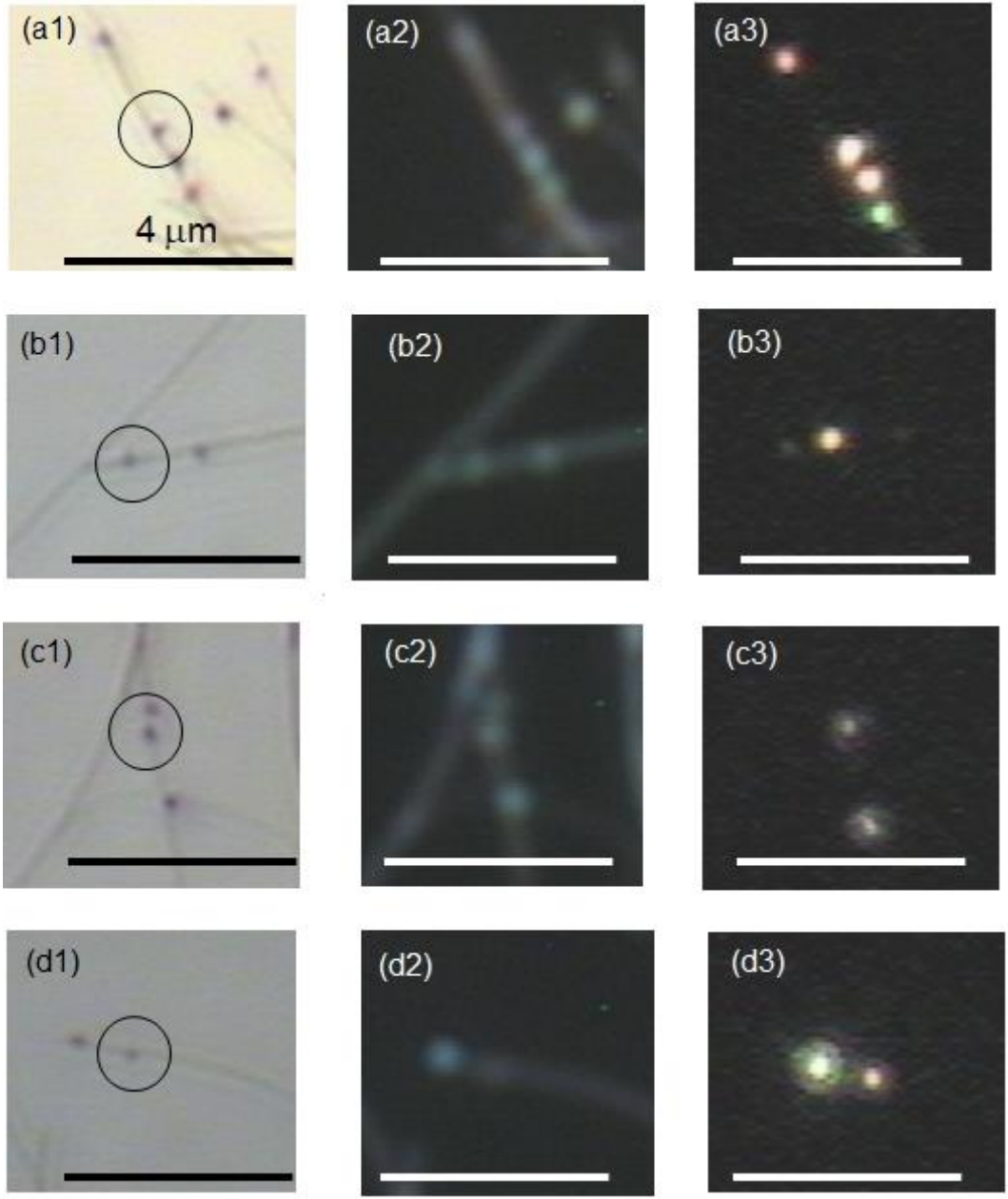

FIG. 9 (a1)–(d1) and (a2)–(d2) Bright- and dark-field images of multiple HSs composed of several NPs on a common NW, respectively. The NPs excited by NIR laser beam are indicated by black open circles in panels (a1)–(d1). (a3)–(d3) SE2PE images of HSs and neighbor HSs on a common NW. All scale bars represent 4 μm.

We found that neighboring NPs on a common NW also emitted SE2PE light. Such propagation phenomena through NWs have been reported for SE1PE light.[29,31,36,37] Figures 9(a1)–9(d1), 9(a2)–9(d2), and 9(a3)–9(d3) show the bright-field, dark-field, and SE2PE images exhibiting propagation phenomena. Two possible propagation mechanisms exist. One is the propagation of SE2PE light supported by the SP through the NWs and emission of SE2PE light at neighboring HSs. In addition, the propagation of NIR light results in the generation and emission of SE2PE light from the neighboring HSs. If the first (or second) mechanism is dominant, the propagation efficiency of visible light (or NIR light) is much higher than that of NIR light (or visible light). Thus, a numerical analysis of the FDTD calculation was used to determine the correct mechanism.

We summarize the observed experimental results as follows.

(1) SE2PE light is generated at the HS between a NP and a NW with CW laser excitation.

(2) The Rayleigh scattering spectra of NPs composed of HSs exhibiting vibrational structures that do not appear in the Rayleigh scattering spectra of isolated NPs.

(3) Large HS-by-HS variations in both SE1PE and SE2PE intensities and intensity noncorrelation between SE1PE and SE2PE.

(4) $PER_{SE2PE} > PER_{SE1PE}$ is commonly observed in HSs.

(5) SE2PE of neighboring HS on a common NW.

We performed numerical calculations to elucidate the above five experimental results, as shown in Figs. 4–9, in which the $F_R$ properties of HSs were induced by EM coupling between the LPs of NPs and SPs of NWs using the FDTD method (EEM-FDM Version 5.1, EEM Co., Ltd., Japan). The complex refractive index of the silver NWs were adopted from a previous study.[20] The effective refractive index of the surrounding medium was set to 1.25 to ensure consistency between the calculated and experimental Rayleigh scattering spectra of the gold NPs.[20] In these calculations, the nonlocal effect, which reduces $F_R$ via Landau damping owing to unscreened surface electrons,[8,9] was not considered because Landau damping does not change the spectral shape of $F_R$ but rather changes its intensity.[38] In the FDTD calculation, we approximated NPs as spherical NPs and NWs as cylindrical NWs with hemispherical ends. However, in a real situation, the shapes of both the NPs and NWs deviate from such approximations. The purpose of the calculation was not to quantitatively reproduce the experimental results but rather to clarify the EM coupling between the LPs of the NPs and SPs of the NWs generating SE2PE.

Figure 10(a1) shows the setup for the FDTD calculation of the NP on the NW in the coordinate system. The beam diameter of the excitation light was 600 nm under these experimental conditions. The excitation polarization directions were set perpendicular

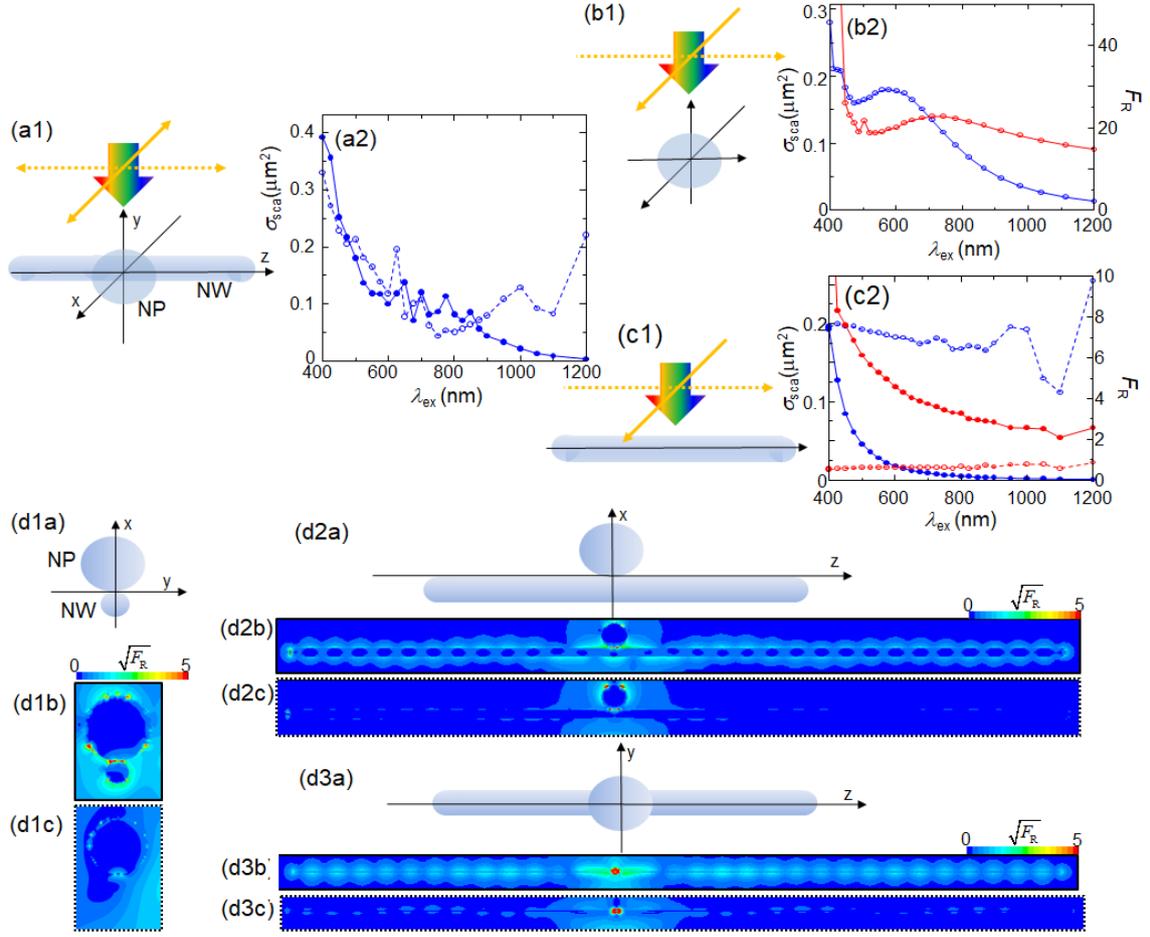

FIG. 10 (a1)–(c1) FDTD calculation setups for $\sqrt{F_R}$ around the system composed of a NP on a NW with $d_g = 0$ nm (a1), isolated NP (b1), and isolated NW (c1) by a narrow-field white light excitation beam with a diameter of 600 nm from the upper side (y-axis). Polarization directions of excitation light are indicated in these panels. (a2) $\sigma_{sca}(\lambda)$ spectra of the NP on NW with the perpendicular (solid blue curve with closed blue circles) and parallel (dashed blue curve with open blue circles) excitation. (b2) and (c2) $\sigma_{sca}(\lambda)$ spectra of the NP and NW, respectively, with the perpendicular (solid blue curves with closed blue circles) and parallel (dashed blue curves with open blue circles) excitation. The maximum $F_R(\lambda_{ex})$ spectra around a NP and NW, respectively, with the perpendicular (solid red curves with closed red circles) and parallel (dashed red curves with open red circles) excitation light. (d1a)–(d3a) FDTD calculation setups for $\sqrt{F_R}$ around the system composed of a NP on a NW showing the x-y-plane at $z = 0$ nm (d1a), x-z plane at $y = 0$ nm (d2a), and y-z plane at $x = 0$ nm (d3a). (d1b)–(d3b) $\sqrt{F_R}$ distributions with perpendicular excitation around the system for the x-y plane at $z = 0$ nm (d1b), x-z plane at $y = 0$ nm (d2b), and y-z plane at $x = 0$ nm (d3b) at $\lambda_{ex} = 525$ nm. (d1c)–(d3c) $\sqrt{F_R}$ distributions with parallel excitation around the system for the x-y plane at $z = 0$ nm (d1c), x-z plane at $y = 0$ nm (d2c), and y-z plane at $x = 0$ nm (d3c) at $\lambda_{ex} = 525$ nm.

and parallel to the long axis of the NW. The $D_{NP}$ and $D_{NW}$ were first set to 200 and 70 nm,

respectively, based on the experimentally measured $D_{NP}$ and $D_{NW}$ in Figs. 2(a2) and 2(b2). The gap distance $d_{gap}$ between the NP and NW was first set to 0 nm, based on the SEM images of the HS in Fig. 5(a1). The length of the NW was 5.5 μm. The amplitude of the incident electric field $|E_{in}|$ was set to be 1.0 V/m. Thus, the value of the calculated amplitude $|E_{loc}|$ corresponds to $|E_{loc}/E_{in}| = \sqrt{F_R}$. Figure 10(a2) shows the spectra of the Rayleigh scattering cross-section $\sigma_{sca}(\lambda_{ex})$ with perpendicular and parallel excitation. The vibrational structures observed in the experimental spectra in Figs. 6(b1)–6(b5) were also reproduced at approximately 600–900 nm. Hence, these vibrational structures are the result of FP interference of the SP, as shown in Fig. 3(a2) along with the EM coupling between the LPs of NP and SPs of NW. Figures 10(b1) and 10(c1) show the calculation setups for the isolated NP and NW in the coordinate systems, respectively. Figures 10(b2) and 10(c2) show the $\sigma_{sca}(\lambda_{ex})$ and $F_R(\lambda_{ex})$ spectra in the polarization directions, respectively. Such vibrational structures do not appear in either the $\sigma_{sca}$ and $F_R$ spectra, confirming that these vibrational structures arise from EM coupling between NP and NW. These results are reasonable because the conversion between light and the SP of NW only occurs at the position of symmetry breaking, such as NPs on NWs, owing to a momentum mismatch.[36]

The EM enhancement at the HS and propagation of the SP through the NW are visualized using FDTD calculations. Figures 10(d1)–10(d3) show the *x-y* ($z = 0$), *x-z* ($y = 0$), and *y-z* ($y = 0$) plane images of $\sqrt{F_R(\lambda_{ex})}$ with perpendicular and parallel excitations with a $\lambda_{ex}$ value of 525 nm. A strong $\sqrt{F_R}$ was observed in the HS. The periodic structures of $\sqrt{F_R}$ in the *x-z* and *y-z* plane images show the propagation of SP waves along the NW from the HS.[29,31,36] The propagation of the SP wave is the origin of the FP interference in Figs. 3(a1), (a2), and 10(a2). The periodic structures appear clearer for perpendicular excitation, indicating that the EM coupling between the LP of the NP and SP of the NW is more efficient for this excitation polarization.

The $\lambda_{ex}$ dependence of the EM enhancement at a HS and propagation of a SP wave through the NW were examined via FDTD calculations. Figures 11(a1) and 11(b1) show the $\sqrt{F_R(\lambda_{ex})}$ spectra along the *z* line ($x, y = 0$) through the HS from a $\lambda_{ex}$ value of 400–1200 nm, visualized as contour maps with perpendicular and parallel excitations, respectively. Figures 11(a2) and 11(b2) show their enlarged images around $-100 < z < 100$ nm. In Figs. 11(a1) and 11(b1), the propagation of the SP wave appears as ripple structures that spread from the HS at approximately $z = 0$, where the largest $\sqrt{F_R}$ is generated, as shown in Figs. 11(a2) and 11(b2). Two spectral regions showing clear ripple structures were observed around the visible (400–900 nm) and NIR (> ~1000 nm) regions in Figs.

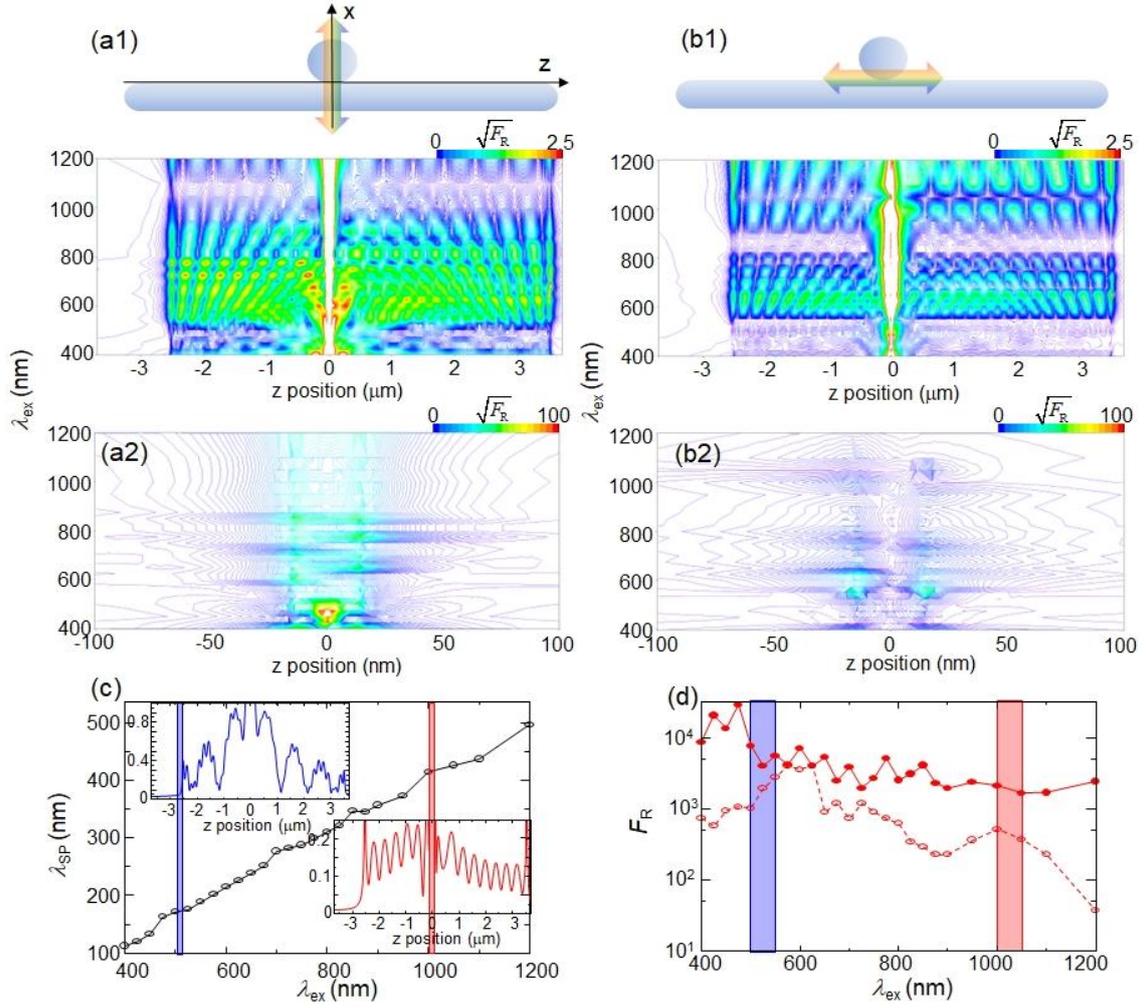

FIG. 11 (a1) and (b1) The $\sqrt{F_R(\lambda_{ex})}$ profiles along the z-axis for $x, y = 0$ nm with $D_{NP}$ and $D_{NW}$ = 200 and 70 nm with $d_g$ = 0 nm at $\lambda_{ex}$ values of 400 to 1200 nm with perpendicular and parallel excitation light, respectively, expressed as contour maps with their illustrations in the upper panels. (a2) and (b2) The $\sqrt{F_R}$ profiles around the z-axis from -100 to 100 nm of (a1) and (b1), expressed as contour maps. (c) Relationship between $\lambda_{SP}$ and $\lambda_{ex}$. The upper and lower panels indicate the $\sqrt{F_R(\lambda_{ex})}$ profiles along the z-axis for $x, y = 0$ nm at $\lambda_{ex}$ values of 500 and 1000 nm, respectively. (d) The $F_R(\lambda_{ex})$ spectra with perpendicular (solid red curve with closed red circles) and parallel (dashed red curve with open red circles) excitation light, respectively. Blue and red boxes indicate the spectral regions used to derive $\overline{F}_R(550\text{nm})$ and $\overline{F}_R(1050\text{nm})$, respectively.

11(a1) and 11(b1), respectively, indicating that there are two types of EM coupling between the LPs of a NP and SPs of the NW. Figure 11(c) shows the relationship between the wavelengths of SP, $\lambda_{SP}$, and $\lambda_{ex}$. The values of $\lambda_{SP}$ were derived from the ripple structures shown in Fig. 11(a1). The coupling between the dipole LP of the NP and the fundamental SP of the NW, which is the lowest-energy radial mode supporting the propagation,[39] is optimized under the condition $D_{NP}$ (200 nm) = $\lambda_{SP}/2$ as illustrated later.[26] Thus, the ripple structures in $\lambda_{ex} < \sim$900 nm ($D_{NP} > \lambda_{SP}/2$) may be mainly induced by EM coupling between the higher-order LPs and SPs. Indeed, the upper $\sqrt{F_R}$ profile at $\lambda_{ex}$ = 500 nm in Fig. 11(c) exhibits a beat structure, which indicates the interference between

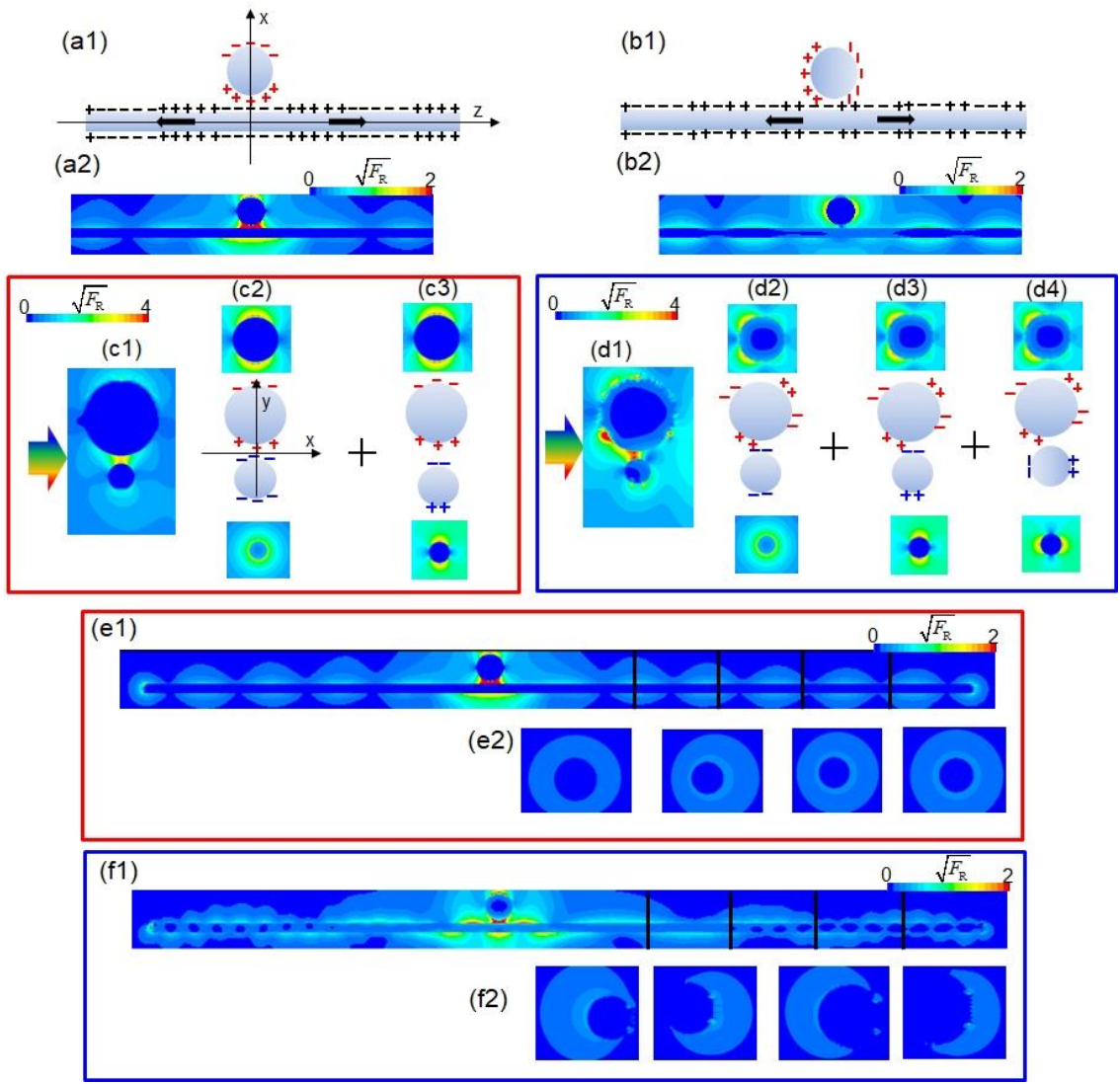

FIG. 12 (a1) and (b1) Charge distributions for the *x-z* plane (*y* = 0 nm) expressing EM coupling between the dipole LP of NP and fundamental SP of NW with perpendicular (a1) and parallel (b1) excitation, respectively. (a2) and (b2) The $\sqrt{F_R}$ distributions for the *x-z* plane at *x* = 0 nm with $D_{NP}$ and $D_{NW}$ = 200 and 70 nm with $d_g$ = 20 nm and $\lambda_{ex}$ = 1200 nm with perpendicular (a2) and parallel (b2) excitation, respectively. (c1) The $\sqrt{F_R}$ distribution for the *x-y* plane at *z* = 0 nm with $d_g$ = 20 nm and $\lambda_{ex}$ = 1200 nm with perpendicular excitation. (c2) and (c3) Charge distributions of the *x-y* plane expressing EM coupling between the dipole LP ($\sqrt{F_R}$ distribution of the upper panel) and fundamental SP ($\sqrt{F_R}$ distribution of the lower panel) and expressing EM coupling between the dipole LP ($\sqrt{F_R}$ distribution of the upper panel) and dipole SP ($\sqrt{F_R}$ distribution of the lower panel) with perpendicular excitation, respectively. (d1) The $\sqrt{F_R}$ distribution for the *x-y* plane at *z* = 0 nm with $d_g$ = 20 nm and $\lambda_{ex}$ = 500 nm with perpendicular excitation. (d2) Charge distribution of the *x-y* plane expressing EM coupling between quadrupole LP ($\sqrt{F_R}$ distribution of upper panel) and fundamental SP ($\sqrt{F_R}$ distribution of the lower panel). (d3) and (d4) Charge distributions of the *x-y* plane expressing EM coupling between quadrupole LP ($\sqrt{F_R}$ distribution of upper panel) and two dipole SPs ($\sqrt{F_R}$ distributions of two lower panels) with perpendicular excitation. The directions of two dipole SPs are orthogonal to each other. (e1) The $\sqrt{F_R}$ distribution for the *x-z* plane at *y* = 0 nm with $d_g$ = 20 nm and $\lambda_{ex}$ = 1200 nm with perpendicular excitation. (e2) $\sqrt{F_R}$ distributions for *x-y* planes at *z* = 1000, 1500, 2000, and 2500 nm, respectively. (f1) $\sqrt{F_R}$ distributions for *x-z* planes at *y* = 0 nm with $d_g$ = 20 nm and $\lambda_{ex}$ = 500 nm with perpendicular excitation. (f2) $\sqrt{F_R}$ distributions for *x-y* planes at *z* = 1000, 1500, 2000, and 2500 nm, respectively.

higher-order SP waves.[40] Such beat structures do not appear under the condition $\lambda_{ex}$ > ~1000 nm ($D_{NP}$ < $\lambda_{SP}$/2) as shown by the lower $\sqrt{F_R(\lambda_{ex})}$ profile at $\lambda_{ex}$ = 1000 nm in Fig. 11(c). Figure 11(d) shows the $F_R(\lambda_{ex})$ spectra at the HS derived from Figs. 11(a2) and 11(b2). The vertical axis represents the logarithmic scale. The vibrational structures in $F_R(\lambda_{ex})$ are similar to those in $\sigma_{sca}(\lambda_{ex})$ in Figs. 10(a2), indicating that both vibrational structures originate from FP interference. The $F_R$ spectrum with the perpendicular excitation was always larger than that with the parallel excitation, indicating that the EM coupling efficiency between the LPs and SPs was better than that for the perpendicular excitation. Furthermore, the intensity ratio between the $F_R$ of the perpendicular excitation and that of the parallel excitation around the NIR region was larger than that around the visible region, supporting the experimentally observed relationship $PER_{SE2PE} > PER_{SE1PE}$ in Fig. 8.

First, we discuss the mechanism of EM coupling between the LPs of a NP and the SPs of a NW in the NIR spectral region. The $d_{gap}$ was set to 20 nm to clarify the structure of the $\sqrt{F_R(\lambda_{ex})}$ around the HSs. Figures 12(a) and 12(b) illustrate the EM coupling between the dipole LP and fundamental SP, which is axially symmetric electrons oscillating parallel to the $z$-axis, for perpendicular and parallel excitation, respectively. The optimal coupling condition of $D_{NP} = \lambda_{SP}/2$ is illustrated in Fig. 12(a1). The strong $\sqrt{F_R}$ at $z = 0$ in the $x$-$z$ plane ($y = 0$) at $\lambda_{ex} = 1200$ nm in Fig. 12(a2) supports this illustration of EM coupling. The optimum coupling condition for the parallel excitation changes to $D_{NP} = \lambda_{SP}/4$, as shown in Fig. 12(b1). The weak $\sqrt{F_R}$ at $z = 0$ in the $x$-$z$ plane ($y = 0$) in Fig. 12(b2) indicates inefficient EM coupling, supporting the discussion in Fig. 11(d) that the $F_R$ for perpendicular excitation is much larger than that for parallel excitation. This difference in the optimum coupling conditions appears as the ripple structures start from $\lambda_{ex} > 850$ nm ($D_{NP} < \lambda_{SP}/4$) for parallel excitation, as shown in Fig. 11(a1). However, they start from $\lambda_{ex} > 1050$ nm ($D_{NP} < \lambda_{SP}/2$) for perpendicular excitation, as shown in Fig. 11(b1).

We discuss the mechanism of EM coupling for the visible and NIR spectral regions, including higher-order LPs and SPs for perpendicular excitation, because the SE2PE intensity of the perpendicular excitation is much larger than that of the parallel excitation, as confirmed in Fig. 11. The $d_{gap}$ was set to be 20 nm to clarify the structure of the $\sqrt{F_R(\lambda_{ex})}$ around the HSs. Figure 12(c1) shows the $\sqrt{F_R}$ distribution in x-z plane ($y = 0$) at $\lambda_{ex} = 1050$ nm. This $\sqrt{F_R}$ distribution is symmetric the y-axis, indicating that the dipole LP of the NP in Fig. 12(c3) contributes to EM coupling in the NIR region. Thus, this $\sqrt{F_R}$ distribution is expressed by the superposition of the EM coupling between the dipole LP and fundamental SP (Fig. 12(c2)) and the EM coupling between the dipole LP and one dipole SPs (Fig. 12 (c3)). We then discuss the EM coupling of the visible regions for $\lambda_{ex}$ of 500–1000 nm in Fig. 11(a1) and $\lambda_{ex}$ of 500–800 nm in Fig. 11(b1). If the $\lambda_{ex}$ region moves to the visible region, then the quadrupole LP of the NP participates in the EM coupling. Figure 12(d1) shows the $\sqrt{F_R}$ distribution in the x-z plane ($y = 0$) at $\lambda_{ex} = 500$ nm. This $\sqrt{F_R}$ distribution is asymmetric against the y-axis, indicating that the quadrupole LP induced by the retardation effect contributes to EM coupling. This quadrupole LP was coupled with the fundamental SPs, as illustrated in Fig. 12(d2). This asymmetric distribution of $\sqrt{F_R}$ causes coupling between the quadrupole LP and two dipole SPs, whose polarization directions are orthogonal to each other, as illustrated in Figs. 12(d3) and 12(d4). The wavenumbers of these two dipole SP waves around the HS should be different because of the presence (Fig. 12(d3)) and absence (Fig. 12(d4)) of the electric field near the HS, where SP waves are tightly confined. Thus, these three SPs are expected to induce the helical propagation of $\sqrt{F_R}$ through the NW.[40] This helical propagation is generated by the coherent interference of the two dipole SPs, whose phases are different from each other. This coherent interference is stretched by the propagating fundamental SP wave.[40] This helical propagation is confirmed as "beat structures" in the distribution of $\sqrt{F_R}$ in the $x$-$z$ plane ($y = 0$). First, we examined the propagation of SP waves in the NIR region. Such beat structures do not appear in the $F_R$ distribution in Fig. 12(e1) with $\lambda_{ex}$ = 1200 nm, indicating that this distribution is determined by the fundamental and dipole SPs. Indeed, the distribution of $\sqrt{F_R}$ in the $x$-$y$ plane in Fig. 12(e2) does not rotate. We then examined the propagation in the visible region. Figure 12(f1) clearly shows the beat structure along the $z$-axis with $\lambda_{ex}$ at 500 nm. Helical propagation was also confirmed by the rotation of the distribution of $\sqrt{F_R}$ in the $x$-$y$ planes, as shown in Fig. 12(f2), supporting the involvement of the quadrupole LP in the EM coupling at

HS in the visible region. The contribution of the quadrupole LP to $F_R$ is important for

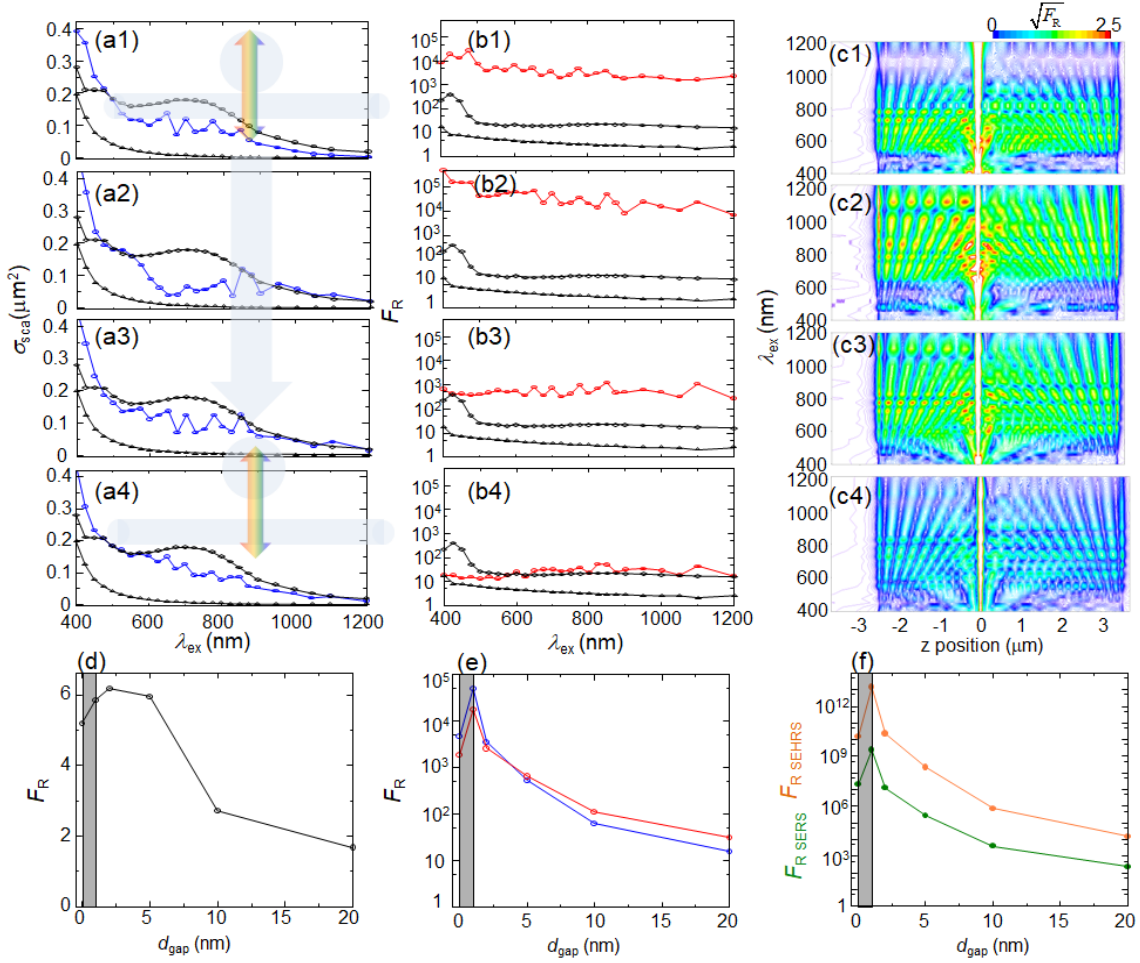

FIG. 13 (a1)–(a4) The $d_g$ dependence of $\sigma_{sca}(\lambda)$ spectra (solid blue curves with open blue circles) of the NP on NW with the perpendicular excitation light with $D_{NP}$ and $D_{NW}$ = 200 and 70 nm, respectively, for $d_g$ = 0, 1, 5, and 20 nm. Black lines with open diamonds and triangles represent the $\sigma_{sca}(\lambda)$ spectra of the isolated NP and NW. (b1)–(b4) The $d_g$ dependence of $F_R(\lambda_{ex})$ spectra (solid red curves with open blue circles) at a HS composed of NP with $D_{NP}$ and $D_{NW}$ = 200 and 70 nm, respectively, for $d_g$ = 0, 1, 5, and 20 nm. Black lines with open diamonds and triangles represent the maximum $F_R(\lambda_{ex})$ spectra of the isolated NP and NW. (c1)–(c4) The $d_g$ dependence of $\sqrt{F_R(\lambda_{ex})}$ profiles along $z$-axis for $x, y$ = 0 nm with $D_{NP}$ and $D_{NW}$ = 200 and 70 nm, respectively, for $d_g$ = 0, 1, 5, and 20 nm with perpendicular excitation expressed as contour maps. (d) Relationship between the maximum $F_R$ in the ripple structures at $z$ ~1 μm in (c1)–(c4) and $d_{gap}$. (e) Relationship between $\overline{F}_R(550\text{nm})$ (blue line with closed circles) (or $\overline{F}_R(1050\text{nm})$ (red line with closed circles)) and $d_{gap}$ at HSs around $z$ = 0 nm. (f) Relationship between $\overline{F}_R(550\text{nm})^2$ (green line with closed circles) (or $\overline{F}_R(1050\text{nm})^2 \overline{F}_R(550\text{nm})$ (ochre line with closed circles)) and $d_{gap}$ at HSs around $z$ = 0 nm.

clarifying the different polarization properties of SE1PE and SE2PE, as discussed in Fig. 8.

We explained that the EM coupling between the LPs of a NP and the SPs of a NW is the origin of both SE2PE at HSs and the vibrational structures in the $\sigma_{sca}(\lambda_{ex})$ spectra. However, there are cases in which HSs showing vibrational structures in $\sigma_{sca}$ hardly generate SE2PE light, as shown in Figs. 6(b5) and 6(c5). The difference between the $d_{gap}$ dependence of $F_R$ and that of the vibrational structures may be the reason for these cases. Thus, we examined the $d_{gap}$ dependences of both $F_R(\lambda_{ex})$ and $\sigma_{sca}(\lambda_{ex})$ spectra, setting $D_{NP}$ and $D_{NW}$ to be 200 and 70 nm, respectively. The polarization direction of the excitation light was set perpendicular to the long axis of the NW. We excluded the discussion for the region with $d_{gap}$ from 0 to 1 nm because the effect of the charge transfer plasmon, which is the coherent oscillation of surface electrons bridging the gap,[5,8,9] is not included in our FDTD calculation. Figures 13(a1)–13(a4) and 13(b1)–13(b4) show the $\sigma_{sca}$ and $F_R$ spectra at the HSs obtained by changing $d_{gap}$ from 0 to 20 nm. With increasing $d_{gap}$, the vibrational structures in $\sigma_{sca}$ remained, but $F_R(\lambda_{ex})$s sharply decreased and almost reached unity at a $d_{gap}$ of 20 nm. These different behaviors between $\sigma_{sca}$ and $F_R$ result in HS showing vibrational structures in $\sigma_{sca}$ but not in SE2PE. Figures 13(c1)–13(c4) show the $\sqrt{F_R}$ spectra along the z-line (x, y = 0) through the HS from $\lambda_{ex}$ values of 400–1200 nm, expressed as contour maps as increasing $d_{gap}$ from 0 to 20 nm. The ripple structures gradually disappeared with an increasing $d_{gap}$, indicating a decrease in the EM coupling energy.

We compared the disappearance of the ripple structures along the NWs in Figs. 13(c1)–13(c4) with the decrease in $F_R(\lambda_{ex})$ in Figs. 13(b1)–13(b4) at the HSs. Figure 13(d) shows the relationship between the maximum $F_R$ in the ripple structures at z ~1 μm and $d_{gap}$. Figures 13(e) shows the relationship between $\bar{F}_R(550\text{nm})$ (or $\bar{F}_R(1050\text{nm})$) and $d_{gap}$ at the HSs around z = 0 nm, where $\bar{F}_R(550\text{nm})$ and $\bar{F}_R(1050\text{nm})$ are the averaged values of $F_R$s for $\lambda_{ex}$ values of 525–575 and 1000–1100 nm, respectively. The $F_R$ in the ripple structures is much less sensitive to the $d_{gap}$ than the $F_R$ at the HSs. This difference indicates that incident light energy that is not tightly confined within the HSs can also be converted into SP waves. That is, the incident light beside the HSs also contributes to the ripple structures, resulting in the loosening of the $d_{gap}$ dependence. Figure 13(f) shows the relationship between $\bar{F}_R(550\text{nm})^2$ (or $\bar{F}_R(1050\text{nm})^2 \bar{F}_R(550\text{nm})$) and the $d_{gap}$ at the HSs, where $\bar{F}_R(550\text{nm})^2$ and $\bar{F}_R(1050\text{nm})^2 \bar{F}_R(550\text{nm})$ represent the EM enhancement factors of SE1PE and SE2PE, respectively. Both $\bar{F}_R(550\text{nm})^2$ and $\bar{F}_R(1050\text{nm})^2 \bar{F}_R(550\text{nm})$ are equally sensitive to $d_{gap}$, indicating that these signals were mainly obtained from HSs with $d_{gap}$ ~ 0 nm, as is supported by the SE1PE and SEM images in Fig. 5(a). However, in these experiments, the HSs exhibiting SE1PE did not exhibit SE2PE, as shown in Figs. 6(c) and 6(d5). These cases suggest that a detectable SE1PE signal can also be generated beside the HSs where the SE2PE signal is too small

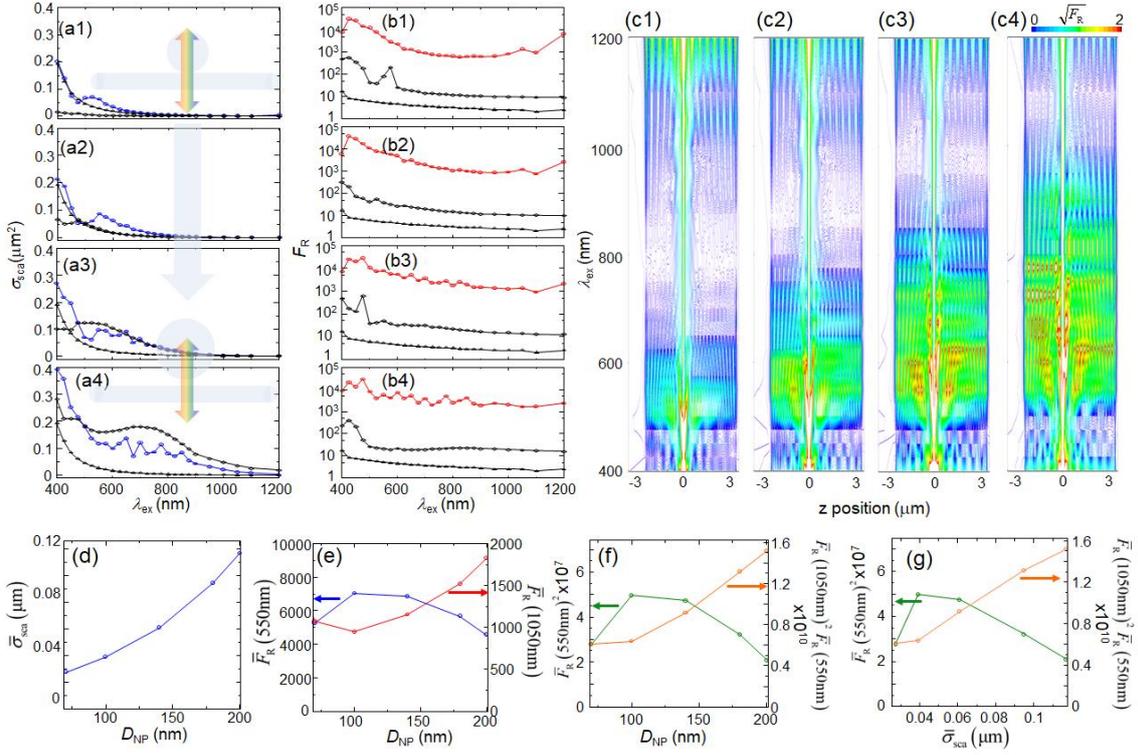

FIG. 14 (a1)–(a4) The $D_{NP}$ dependence of $\sigma_{sca}$ spectra (solid blue curves with open blue circles) of the NP on NW with perpendicular excitation light with $d_g$ and $D_{NW}$ = 0 and 70 nm, respectively, for $D_{NP}$ = 70, 100, 140, and 200 nm. Black lines with open diamonds and triangles represent the $\sigma_{sca}$ spectra of the isolated NP and NW. (b1)–(b4) The $D_{NP}$ dependence of $F_R$ spectra (solid red curves with open red circles) at HSs composed of NP with $d_g$ and $D_{NW}$ = 0 and 70 nm, respectively, for $D_{NP}$ = 70, 100, 140, and 200 nm. Black lines with closed diamonds and triangles represent the maximum $F_R$ spectra of the isolated NP and NW. (c1)–(c4) The $D_{NP}$ dependence of $\sqrt{F_R}$ profiles along the z-axis for $x, y$ = 0 nm with $d_g$ and $D_{NW}$ = 0 and 70 nm, respectively, for $D_{NP}$ = 70, 100, 140, and 200 nm with perpendicular excitation expressed as contour maps. (d) Relationship between the maximum $\overline{\sigma}_{sca}$ of the NP on NW and $D_{NP}$. (e) Relationship between $\overline{F}_R(550\mathrm{nm})$ (blue line with closed circles) (or $\overline{F}_R(1050\mathrm{nm})$ (red line with closed circles)) at HSs around $z$ = 0 nm and $D_{NP}$. (f) Relationship between $\overline{F}_R(550\mathrm{nm})^2$ (green line with closed circles) (or $\overline{F}_R(1050\mathrm{nm})^2 \overline{F}_R(550\mathrm{nm})$ (ochre line with closed circles)) at HSs around $z$ = 0 nm and $D_{NP}$. (g) Relationship between $\overline{F}_R(550\mathrm{nm})^2$ (green line with closed circles) (or $\overline{F}_R(1050\mathrm{nm})^2 \overline{F}_R(550\mathrm{nm})$ (ochre line with closed circles)) at HSs around $z$ = 0 nm and $\overline{\sigma}_{sca}$.

to detect. This suggestion is consistent with the finding that SE1PE (SE2PE) is mainly composed of SEF (SEHRS) and not SERS (two-photon SEF), as discussed in Fig. 6. Furthermore, the $\bar{F}_R(1050\text{nm})^2 \bar{F}_R(550\text{nm})$ sensitivity to $d_{\text{gap}}$, showing intensity changes of $10^3$ times within 0–2 nm of $d_{\text{gap}}$, can be the main origin of the large HS-by-HS variations in SE2PE intensities in Fig. 7(c). The lack of correlation between the SE1PE and SE2PE intensities in Fig. 7(d) can also be explained by the presence or absence of the contribution of signals other than the HSs for SE1PE and SE2PE, respectively.

We calculated the $D_{\text{NP}}$ dependence of $F_R(\lambda_{\text{ex}})$ to investigate the large HS-by-HS variations in the SE1PE and SE2PE intensities and the lack of correlation between the SE1PE and SE2PE intensities, as shown in Fig. 7. Figures 14(a1)–14(a4) and 14(b1)–14(b4) show the $\sigma_{\text{sca}}(\lambda_{\text{ex}})$ and $F_R(\lambda_{\text{ex}})$ spectra of $D_{\text{NP}}$ from 70–200 nm under perpendicular excitation. These $\sigma_{\text{sca}}$ spectral maxima exhibit the red-shifts and subsequent broadening increasing $D_{\text{NP}}$, showing the vibrational structures, which are the effects of the FP interference of the NWs, as shown in Fig. 10(a2). The maximum values of $\sigma_{\text{sca}}$ for $\lambda_{\text{ex}} >$ 500 nm in Figs. 14(a3)–14(a4) are always smaller than the $\sigma_{\text{sca}}$ values of isolated NPs, indicating light energy transfer from radiative LPs of NP to nonradiative SPs of NW via EM coupling. The EM coupling also confines the incident light within the HSs. Thus, the values of $F_R$ at the HSs are always much larger than the $F_R$ values of both the isolated

NPs and NWs, as shown in Figs. 14(b1)–14(b4). The $F_R$ spectra also exhibited broadening with increasing $D_{NP}$ with indicating vibrational structures caused by FP interference. The broadening of the $F_R(\lambda_{ex})$ spectra is attributable to the red-shifts of the quadrupole LP resonance, which is coupled with both the fundamental and dipole SPs, as shown in Figs. 12(d1)–12(d4). Figures 14(c1)–14(c5) show the $\sqrt{F_R(\lambda_{ex})}$ spectra along the z-line (x, y = 0) through HSs from $\lambda_{ex}$ of 400–1200 nm, expressed as contour maps as increasing $D_{NP}$ from 50 to 200 nm. The ripple structures in both the visible and NIR regions exhibited redshifts with increasing $D_{NP}$. The red-shifts in $\sqrt{F_R}$ can be explained by the red-shifts in the coupled resonance maxima between the quadrupole LP and SPs in the visible region and in the coupled resonance maxima between the dipole LP and SPs in the NIR region.

The relationship between Rayleigh scattering intensities and both SE1PE and SE2PE intensities, as shown in Fig. 7(b), was examined by changing $D_{NP}$. The Rayleigh scattering intensities were reproduced as $\sqrt{F_R}$, which represents the average values of $\sigma_{sca}(\lambda_{ex})$ with $\lambda_{ex}$ of 500–800 nm. The relationship between $\sqrt{F_R}$ and $D_{NP}$ in Fig. 14(d) shows that the $\sqrt{F_R}$ simply increases with an increasing $D_{NP}$. Figure 14(e) shows the relationship between $\bar{F}_R(550\text{nm})$ ($\bar{F}_R(1050\text{nm})$) and $D_{NP}$. The $\bar{F}_R(550\text{nm})$ achieves its maximum at a $D_{NP}$ of approximately 100 nm, and then decreases. The $\bar{F}_R(1050\text{nm})$ achieves its minimum at a $D_{NP}$ of approximately 100 nm and then increases. These behaviors of $\bar{F}_R(550\text{nm})$ and $\bar{F}_R(1050\text{nm})$ are determined by their spectral overlapping with the coupled resonances at HSs, as shown in Figs. 14(c1)–14(c4). That is, when the coupled resonance maxima get close to (or leave) 550 nm or 1050 nm, $\bar{F}_R(550\text{nm})$ or $\bar{F}_R(1050\text{nm})$ increases (or decrease). Figure 14(f) shows the relationship between $\bar{F}_R(550\text{nm})^2$ ($\bar{F}_R(1050\text{nm})^2 \bar{F}_R(550\text{nm})$) and $D_{\text{NP}}$. To compare these calculated tendencies against $D_{\text{NP}}$ with the experimental results, the $D_{\text{NP}}$ dependencies change into the $\sqrt{F_R}$ dependencies as shown in Fig. 14(d). Figure 14(g) shows the relationship between $\sqrt{F_R}$ and $\bar{F}_R(550\text{nm})^2$ (or $\bar{F}_R(1050\text{nm})^2 \bar{F}_R(550\text{nm})$). The calculated $\sqrt{F_R}$ dependencies do not explain the large HS-by-HS variations in SE1PE and SE2PE as in Fig. 7(b). Thus, the experimentally observed large HS-by-HS variations may not be attributable to the variations in $D_{\text{NP}}$. The uncorrelation between SE1PE and SE2PE as in Fig. 7(c) is consistent with the results shown in Fig. 14(g), indicating that the uncorrelation is related to the absence of $\bar{F}_R(1050\text{nm})$ in SE1PE.

We examined the $D_{\text{NW}}$ dependence of $F_R(\lambda_{\text{ex}})$ to determine the large HS-by-HS distribution shown in Fig. 7. The excitation polarization and $D_{\text{NP}}$ were set to be perpendicular to the long axis of the NW and 200 nm, respectively. Figures 15(a1)–15(a4) and 15(b1)–15(b4) show the $\sigma_{\text{sca}}(\lambda_{\text{ex}})$ and $F_R(\lambda_{\text{ex}})$ of $D_{\text{NW}}$ from 70 to 200 nm. The $\sigma_{\text{sca}}$ spectral intensities increased, and the vibrational structures became unclear with an

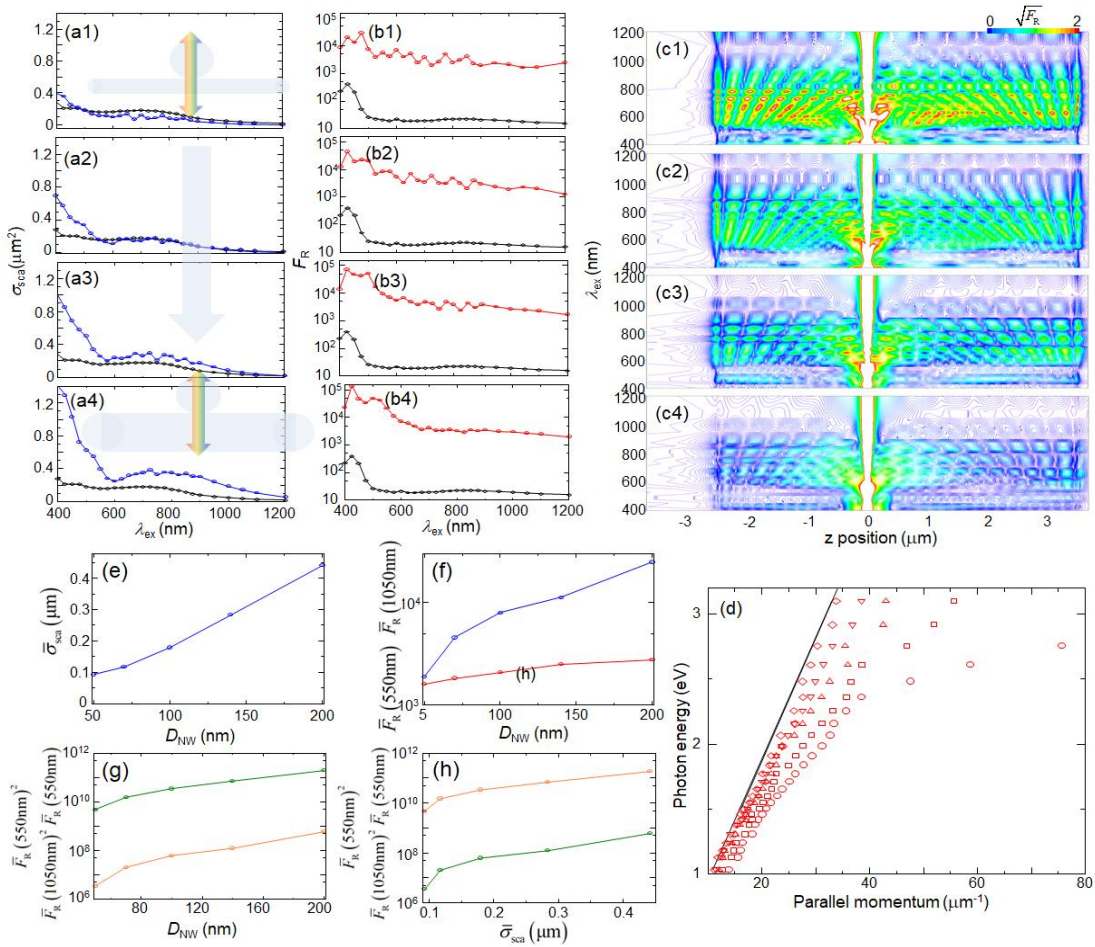

FIG. 15 (a1)–(a4) The $D_{NW}$ dependence of $\sigma_{sca}(\lambda)$ spectra (solid blue curves with open blue circles) of the NP on NW with perpendicular excitation light with $d_g$ and $D_{NP} = 0$ and 200 nm, respectively, for $D_{NW}$ = 70, 100, 140, and 200 nm. Black lines with closed diamonds and triangles represent the $\sigma_{sca}(\lambda)$ spectra of the isolated NP and NW, respectively. (b1)–(b4) The $D_{NW}$ dependence of $F_R(\lambda_{ex})$ spectra (solid red curves with open red circles) at HS composed of NP with $d_g$ and $D_{NP} = 0$ and 200 nm, respectively, for $D_{NW}$ = 70, 100, 140, and 200 nm. Black lines with closed diamonds and triangles represent the maximum $F_R$ spectra of the isolated NP and NW, respectively. (c1)–(c4) The $D_{NW}$ dependence of $\sqrt{F_R}$ profiles along the z-axis for x, y = 0 nm with $d_g$ and $D_{NP} = 0$ and 200 nm, respectively, for $D_{NW}$ = 70, 100, 140, and 200 nm with perpendicular excitation expressed via contour maps. (d) The $D_{NW}$ dependence of dispersion relationships between the wavenumber and energy of the fundamental SP for $D_{NW}$ = 50 (○), 70 (□), 100 (△), 140 (▽), and 200 (◇) nm. (e) Relationship between the maximum $\sqrt{F_R}$ of the NP on NW and $D_{NW}$. (f) Relationship between $\bar{F}_R(550\text{nm})$ (blue line with closed circles) (or $\bar{F}_R(1050\text{nm})$ (red line with closed circles)) at HSs around $z = 0$ nm and $D_{NW}$. (g) Relationship between $\bar{F}_R(550\text{nm})^2$ (green line with closed circles) (or $\bar{F}_R(1050\text{nm})^2 \bar{F}_R(550\text{nm})$ (ochre line with closed circles)) at HSs around $z = 0$ nm and $D_{NW}$. (h) Relationship between $\bar{F}_R(550\text{nm})^2$ (green line with closed circles) (or $\bar{F}_R(1050\text{nm})^2 \bar{F}_R(550\text{nm})$ (ocher line with closed circles)) at HSs around z = 0 nm and $\sqrt{F_R}$.

increasing $D_{NW}$, as shown in Fig. 15(a4). The values of $F_R$ at the HSs were always much larger than the $F_R$ of the isolated NPs. The vibrational structures in the $F_R$ spectra became unclear as increasing $D_{NW}$, as shown in Fig. 15(a4). This ambiguity is considered to be induced by the increase in the period of the vibrational structures of FP interference because the field confinement by SPs is loosened with increasing $D_{NW}$. Thus, the $D_{NW}$ dependence of the ripple structures was verified using $\sqrt{F_R(\lambda_{ex})}$ spectra. Figures 15(c1)–15(c4) show the $\sqrt{F_R}$ spectra along the $z$-line ($x, y = 0$) through HSs from $\lambda_{ex}$ values of 400–1200 nm, expressed as contour maps by increasing the $D_{NW}$ from 70 to 200 nm. The period of the vibrational structures clearly increases with increasing $D_{NW}$, supporting the idea that the loosening of field confinement arises from the spectral features in Figs. 15(a) and 15(b). Figures 15(d) shows the $D_{NW}$ dependence of the dispersion relationships between the wavenumbers and energies of the fundamental SP derived from Figs. 15(c1)–15(c4). Decreases in the wavenumbers of the SP are clearly observed with increasing $D_{NW}$, thus confirming our discussion of the loosening of field confinement.

The $D_{NW}$ dependences of the relationships between the Rayleigh scattering intensities and both SE1PE and SE2PE intensities in Fig. 7(b) were examined. Figure 15(e) shows the relationship between $D_{NW}$ and $\bar{\sigma}_{sca}$. The $\bar{\sigma}_{sca}$ increased monotonously with increasing $D_{NW}$. Figure 15(f) shows the relationship between the $D_{NW}$ and $\bar{F}_R(550\text{nm})$ ($\bar{F}_R(1050\text{nm})$). Both $\bar{F}_R(550\text{nm})$ and $\bar{F}_R(1050\text{nm})$ increase with an increasing $D_{NW}$.

These properties may be induced by approaching the dipole LP (not SP) resonance of NW

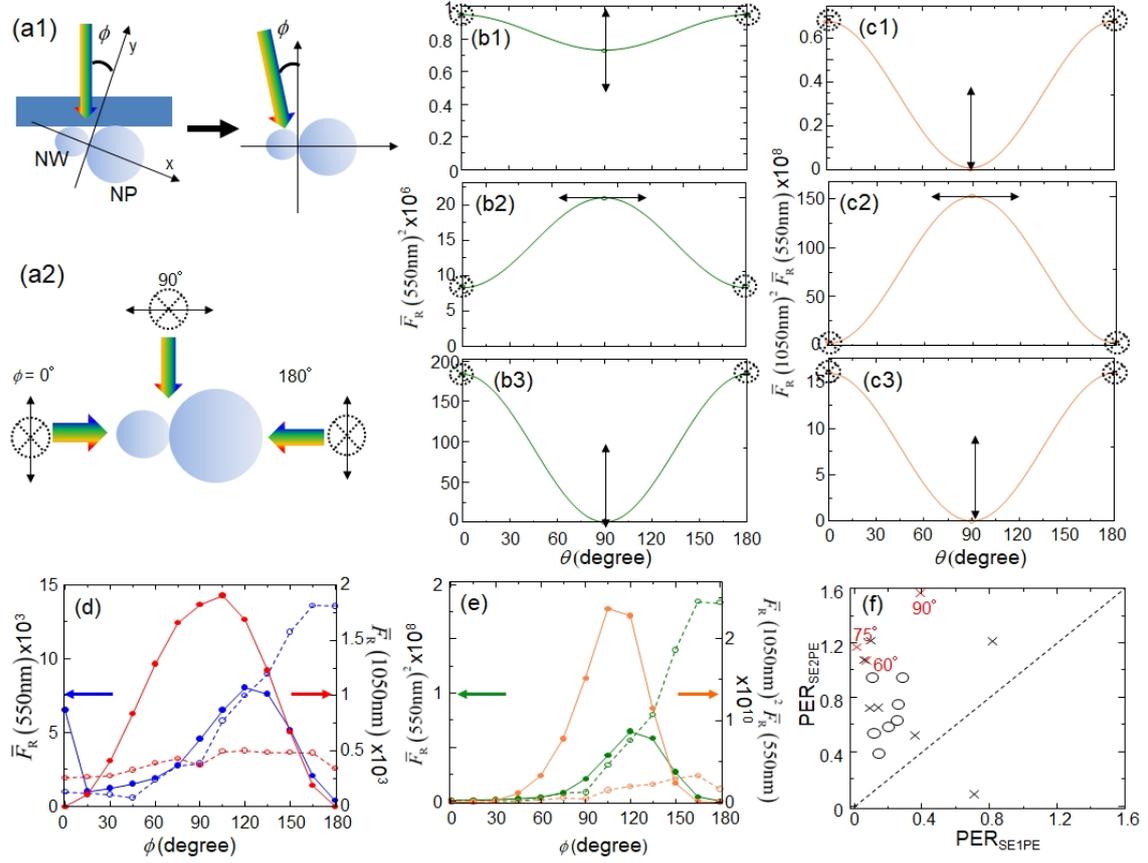

FIG. 16 (a1) NP and NW on a common glass plane (left panel) corresponds to the tilted excitation beam angle (right panel). Here, $D_{NP}$ and $D_{NW}$ = 200 and 70 nm, respectively, for $d_g$ = 0 nm. (a2) Definitions of $\phi$ for 0°, 90°, and 180°, respectively. Arrows and circles with crosses indicate perpendicular and parallel excitation, respectively. (b1)–(b3) and (c1)–(c3) The $\theta$ dependencies of $\bar{F}_R(550\text{nm})^2$ and $\bar{F}_R(1050\text{nm})^2 \bar{F}_R(550\text{nm})$, respectively, with $\phi$ values of 0° (b1, c1), 90° (b2, c2), and 180° (b3, c3). (d) The $\phi$ dependence of $\sqrt{F_R(\lambda_{ex})}$ with perpendicular (blue lines with closed circles) and parallel excitation (blue lines with open circles). The $\phi$ dependence of $\bar{F}_R(1050\text{nm})$ with perpendicular (red lines with closed circles) and parallel excitation (red lines with open circles). (e) The $\phi$ dependence of $\bar{F}_R(550\text{nm})^2$ with perpendicular (green lines with closed circles) and parallel excitation (green lines with open circles). The $\phi$ dependence of $\bar{F}_R(1050\text{nm})^2 \bar{F}_R(550\text{nm})$ with perpendicular (ochre lines with closed circles) and parallel excitation (ochre lines with open circles). (f) Experimental (open circles) and calculated (diagonal crosses) relationships between $PER_{SE1PE}$ and $PER_{SE1PE}$. The points corresponding to $\phi$ for 60°, 75°, and 90° are indicated in this panel.

to LP resonances of NP, resulting in efficient EM coupling. Red-shifts in the dipole LP resonance maxima of NW are observed around $\lambda_{ex}$ values of 400 nm in Figs. 15(a1)–15(a4) with increasing $D_{NW}$. Figure 15(g) shows the relationship between the $D_{NW}$ and $\bar{F}_R(550\text{nm})^2$ ( $\bar{F}_R(1050\text{nm})^2 \bar{F}_R(550\text{nm})$ ). Both $\bar{F}_R(550\text{nm})^2$ and $\bar{F}_R(1050\text{nm})^2 \bar{F}_R(550\text{nm})$ increase monotonically with an increasing $D_{NW}$. To compare these calculated tendencies with the experimental results, the $D_{NW}$ dependencies in Fig. 14(g) were changed into $\bar{\sigma}_{sca}$ dependencies. Figure 15(h) shows the relationship between $\bar{\sigma}_{sca}$ and $\bar{F}_R(550\text{nm})^2$ ( $\bar{F}_R(1050\text{nm})^2 \bar{F}_R(550\text{nm})$ ). The calculated results shown in Fig. 15(h) do not reproduce the experimental tendencies shown in Fig. 7(b). Thus, the large HS-by-HS variations in both SE1PE and SE2PE cannot be explained by the variations in $D_{NW}$. However, the various vibrational structures in $\sigma_{sca}(\lambda_{ex})$ in Figs. 6(b1)–(b5) may be related to the $D_{NW}$ dependences of $\lambda_{SP}$, as shown in Figs. 15(c1)–(c4).

The experimental polarization dependencies of the SE1PE and SE2PE intensities shown in Fig. 8 were evaluated via FDTD calculations using $\bar{F}_R(550\text{nm})^2$ and $\bar{F}_R(1050\text{nm})^2 \bar{F}_R(550\text{nm})$, respectively. The tendency of $PER_{SE1PE}$ is always smaller than $PER_{SE2PE}$, as shown in Fig. 8(b). The FDTD calculations in Figs. 13–15 assume that the NP and NW are aligned in parallel and that the incident light beam comes from the upper side, as shown in Fig. 10(a1). However, in reality, both the NP and NW are placed on a common glass plane, as shown in Fig. 16(a1). Thus, the effective direction of the incident light beam should be tilted within a tilt angle ($\phi$) range of approximately 70° < $\phi$ < 90°, as shown in Fig. 16(a1). Thus, to evaluate the variations in $\phi$, we examined the $\theta$ dependence of $\bar{F}_R(550\text{nm})^2$ and $\bar{F}_R(1050\text{nm})^2 \bar{F}_R(550\text{nm})$ by changing $\phi$ from 0° to 180°, as shown in Fig. 16(a2). The $D_{NW}$ and $D_{NP}$ values were set to 70 and 200 nm, respectively, with $d_g$ = 0 nm. Figures 16(a1)–16(a3) and 16(b1)–16(b3) show the $\theta$ dependencies of $\bar{F}_R(550\text{nm})^2$ and $\bar{F}_R(1050\text{nm})^2 \bar{F}_R(550\text{nm})$, respectively, with $\phi$ values of 0°, 90°, and 180°. Both $\theta$ dependencies changed their phases by 180° when $\phi$ changed from 0° to 90°. The $\theta$ dependencies further changed by 180° when $\phi$ changed from 90° to 180°. Regarding the range of $\phi$ approximately 70° < $\phi$ < 90°, Figs. 16(b2) and 16(c2) showing PER$_{SE1PE}$ < PER$_{SE1PE}$ reasonably reproduce the experimental results in Fig. 8(a). The EM coupling discussed in Figs. 12(d1)–12(d4) indicates that the complicated electric fields induced by the quadrupole LP of the NP around the $\bar{F}_R(550\text{nm})$ region make the polarization dependence of SE1PE unclear, resulting in PER$_{SE1PE}$ < PER$_{SE2PE}$.

The $\phi$ dependencies of $\theta$ dependence of $\bar{F}_R(550\text{nm})$ and $\bar{F}_R(1050\text{nm})$ were investigated in detail. Figure 16(d) shows the $\phi$ dependencies on the excitation light polarized perpendicularly ($\theta$=90°) and parallel ($\theta$=0, 180°) to the long axis of the NW.

For perpendicular excitation, the $\bar{F}_R(550\text{nm})$ exhibits its maximum and minimum at $\phi$ values of approximately 120° and 180°, respectively. However, $\bar{F}_R(550\text{nm})$ exhibits its extreme values at $\phi$ values of approximately 100° and 0° (and 180°). For parallel excitation, the $\bar{F}_R(550\text{nm})$ exhibits its maximum and minimum at $\phi$ values of 180° and 0°, respectively; however, $\bar{F}_R(1050\text{nm})$ does not exhibit a clear $\phi$ dependence. In short, these $\phi$ dependences are significantly different from each other. Here, $\bar{F}_R(550\text{nm})$ was generated by EM coupling involving the quadrupole LPs of the NPs, as shown in Fig. 12(d). In contrast, $\bar{F}_R(1050\text{nm})$ is generated by EM coupling involving the dipole LPs of a NP, as in Fig. 12(c). Thus, the difference in $\phi$ dependence between $\bar{F}_R(550\text{nm})$ and $\bar{F}_R(1050\text{nm})$ is induced by the presence or absence of the contribution of the quadrupole LP to the EM coupling. That is, the complex electric field of the quadrupole LP in Fig. 12(d1) makes their $\phi$ dependencies deviate from $\cos^2(\theta)$. Figure 16(e) shows the $\phi$ dependencies of $\bar{F}_R(1050\text{nm})$ and $\bar{F}_R(1050\text{nm})^2 \bar{F}_R(550\text{nm})$ derived from Fig. 16(d) on the perpendicular and parallel excitations. The relationship between $PER_{SE1PE} < PER_{SE2PE}$ in the experiments shown in Fig. 8 was reasonably reproduced within the range 70° < $\phi$ < 90°. To check Fig. 8(b), we derived $PER_{SE1PE}$ and $PER_{SE2PE}$ from Fig. 16(e). Figure 16(f) shows that the calculated $PER_{SE1PE}$ is always smaller than the calculated $PER_{SE1PE}$ for $\phi$ values of 65°, 75°, and 90°, reproducing the experimental results well.

Thus, we conclude that the relationship $PER_{SE1PE} < PER_{SE2PE}$ is induced by the contribution of the quadrupole LP to the EM coupling. Additionally, we consider that the $\phi$ dependencies can be minor origins of the large HS-by-HS variations in SE2PE, as shown in Fig. 7.

We evaluated the propagation phenomena of SE2PE from one HS to the next HS, as shown in Fig. 9, using FDTD calculations. There are two possible mechanisms: the propagation of SE2PE light and the propagation of NIR light through the NW. These two mechanisms were examined by using systems comprising two NPs on a common NW, as shown in Fig. 17(a). The $D_{NW}$ and $D_{NP}$ of both the NPs were set to 70 and 200 nm, respectively, with $d_g = 0$ nm. Figures 17(b1) and 17(b2) show the $x$-$z$ plane images ($y = 0$) of the $\sqrt{F_R}$ with the excitation light polarized perpendicular and parallel to the long axis of the NW at a $\lambda_{ex}$ value of 550 nm, respectively. Both exhibit the propagation of SP waves, and the propagation was more clearly observed for the perpendicular excitation, reflecting efficient EM coupling, as shown in Fig. 12. We verified the propagation of SP waves using ripple structures in the $\sqrt{F_R}$ spectra. Figures 17(c1) and 17(c2) show the $\sqrt{F_R}$ spectra along a $z$-line ($x, y = 0$) through the HS from $\lambda_{ex}$ values of 400–1200 nm, expressed via contour maps with perpendicular and parallel excitations, respectively. Both exhibited a SP wave propagating from the first NP through the NW and contacting the second NP, resulting in the second HS. The ripple structures between the first and second NPs are similar to those shown in Figs. 11(a1) and 11(b1). However, the ripple structures weakened after the second NP owing to the radiative loss of energy of the SP waves at the second HS.

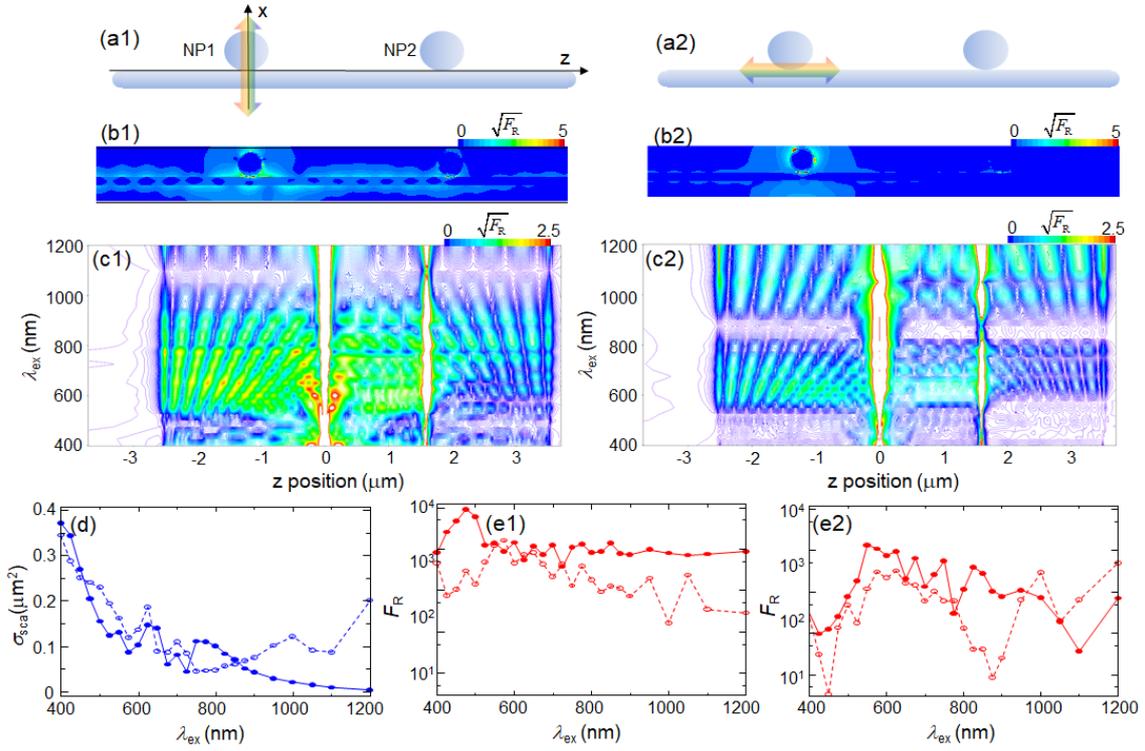

FIG. 17 (a1) and (a2) The system composed of NP1 and NP2 with $D_{NPs}$ = 200 nm on a common NW for $D_{NW}$ =70 nm with both $d_g$s = 0 nm with perpendicular (a1) and parallel (a2) excitations of NP1. (b1) and (b2) The $\sqrt{F_R}$ distributions for the y-z planes at x = 0 nm at a $\lambda_{ex}$ value of 550 nm with perpendicular (b1) and parallel (b2) excitations of NP1. (c1) and (c2) The $\sqrt{F_R(\lambda_{ex})}$ profiles along the z-axis for x, y = 0 nm from $\lambda_{ex}$ values of 400–1200 nm with perpendicular (c1) and parallel (c1) excitations, respectively, expressed via contour maps. (d) The $\sigma_{sca}(\lambda)$ spectra of NP1 on NW with perpendicular (solid blue curve with closed blue circles) and parallel (dashed blue curve with open blue circles) excitation. (e1) and (e2) The $F_R(\lambda_{ex})$ spectra with perpendicular (solid red curve with closed red circles) and parallel (dashed red curve with open red circles) excitations, respectively, of the first (e1) and second (e2) HSs, respectively.

Figure 17(d) shows the $\sigma_{sca}(\lambda_{ex})$ spectra under perpendicular and parallel excitations.

These spectra are almost the same as those without the second NP, as shown in Fig. 10(a2), indicating that the effect of the reflected SP waves from the second NP is negligible. Figures 17(e1) and 17(e2) show $F_R(\lambda_{ex})$ with perpendicular and parallel excitations at the first and second HSs, respectively. The $\bar{F}_R(550\text{nm})$ of the first HS is approximately two times larger than that of the second HS, and the $\bar{F}_R(1050\text{nm})$ of the first HS is five times greater than that of the second HS. Thus, if the observed SE2PE in the second HS is the propagated SE2PE light from the first HS, the SE2PE intensity of the second HS is expected to be half of the SE2PE intensity of the first HS. However, if the observed SE2PE at the second HS is generated by NIR light propagating from the first HS, the SE2PE intensity of the second HS is only 4% of the SE2PE intensity of the first HS. Thus, the observed SE2PE signals from the second HS were mainly the SE2PE signals coming from the first HS.

We examined the dependence of the SE2PE intensity of the second HS on the propagation distance. Figures 18(a1)–18(a3) illustrate the increase in the distance between the first (NP1) and second (NP2) NPs from 500 to 2500 nm. The $D_{NW}$ and $D_{NP}$ of both the NPs were set to 70 and 200 nm, respectively, with $d_g = 0$ nm. Figures 18(b1) and 18(b2) show the $\sqrt{F_R(\lambda_{ex})}$ profiles along the z-line (x, y = 0) through a HS at $\lambda_{ex}$

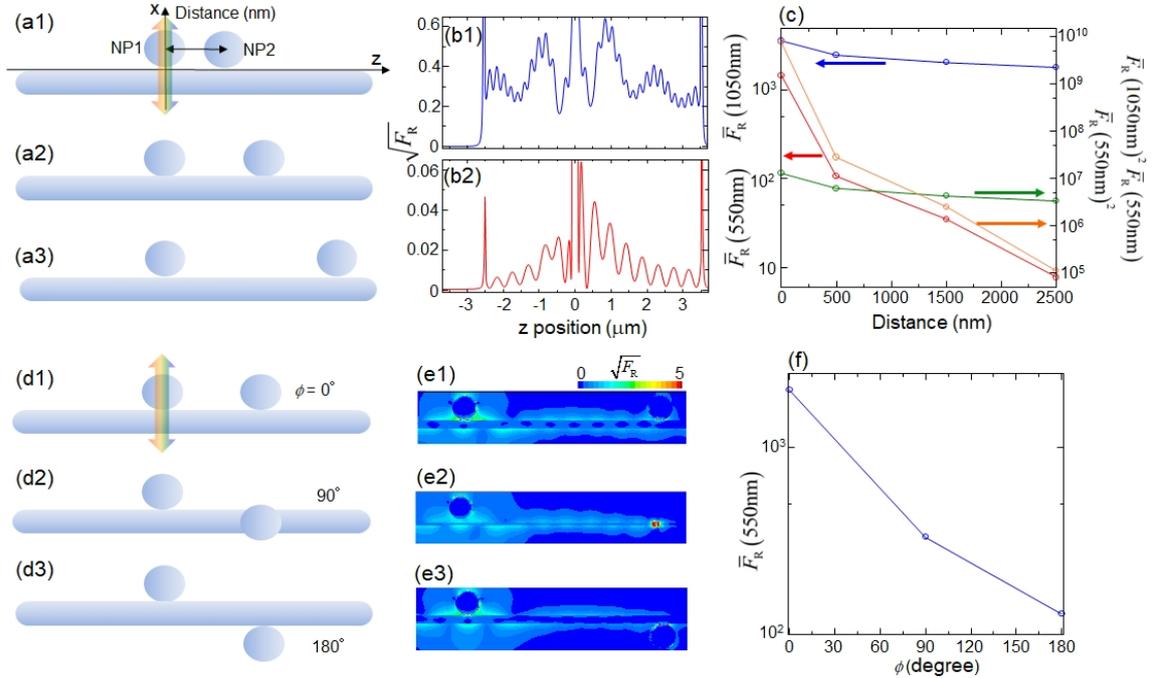

FIG. 18 (a1)–(a3) The system composed of NP1 and NP2 with $D_{NP} = 200$ nm on a common NW with $D_{NW} = 70$ nm for $d_g = 0$ nm with perpendicular excitation of NP1. The distances between NP1 and NP2 were 500, 1500, 2500 nm, respectively. (b1) and (b2) The $\sqrt{F_R(\lambda_{ex})}$ profiles along the z-axis for x, y = 0 nm with $\lambda_{ex}$ values of 550 and 1050 nm with perpendicular excitation light. (c) The NP1–NP2 distance dependencies of $\overline{F}_R(550\text{nm})$ (blue lines with open circles), $\overline{F}_R(1050\text{nm})$ (red lines with open circles), $\overline{F}_R(550\text{nm})^2$ (green lines with open circles), and $\overline{F}_R(1050\text{nm})^2 \overline{F}_R(550\text{nm})$ (ochre lines with open circles). (d1)–(d3) The system composed of two NPs with $D_{NP} = 200$ nm on a common NW $D_{NW} = 70$ nm with both $d_g = 0$ nm with perpendicular excitation of NP1 with the $\phi$ of NP2 set at 0°, 90°, and 180° against NP1. (e1)–(e3) The $\sqrt{F_R}$ distribution for the y-z plane at x = 0 nm with $\phi$ of 0° (e1), x = 35 nm with $\phi$ of 90° (e2), and at x = 0 nm with $\phi$ of 180° (e3). (c) The $\phi$ dependence of $\overline{F}_R(550\text{nm})$ (blue lines with open circles).

values of 550 and 1050 nm, respectively, with perpendicular excitation. The $z$-line profile for $\lambda_{ex}$ of 1050 nm exhibited faster decay than that for $\lambda_{ex}$ of 550 nm. The propagation at $\lambda_{ex}$ of 1050 nm is primarily supported by the fundamental SP, in which the electric field penetrates the metal deeper than does the electric field supported by dipole SP,[39] resulting in a greater Ohmic loss than that of dipole SP. Figure 18(c) shows the propagation distance dependencies of $\bar{F}_R(550\text{nm})$, $\bar{F}_R(1050\text{nm})$, $\bar{F}_R(550\text{nm})^2$, and $\bar{F}_R(1050\text{nm})^2 \bar{F}_R(550\text{nm})$. Here, $\bar{F}_R(550\text{nm})$ and $\bar{F}_R(550\text{nm})^2$ exhibited smaller propagation losses than $\bar{F}_R(1050\text{nm})$ and $\bar{F}_R(1050\text{nm})^2 \bar{F}_R(550\text{nm})$, thus confirming that the observed SE2PE from the second HS was propagating SE2PE light from the first HS.

The considerable variations in the SE2PE intensities at the second HSs are observed in Fig. 9, similar to the intensity variations at the first HS in Fig. 7. We discuss the origins of the variation for the first HSs as the $d_{gap}$ dependence in Fig. 13 and $\phi$ dependence in Fig. 16. The $d_{gap}$ dependence is the origin of variations in the second HSs. Thus, we examined the $\phi$ dependence of the SE2PE intensities for the second HSs. Figures 18(d1)–18(d3) illustrate that NP2 has three values of $\phi$ against NP1, set at 0°, 90°, and 180°. The $D_{NW}$ and $D_{NP}$ of both the NPs were set to 70 and 200 nm, respectively, with $d_g = 0$ nm. The SE2PE of the second HS was the SE2PE light propagating from the first HS. Thus, we calculated the $\phi$ dependence of the propagation of SP waves at $\lambda_{ex}$ of 550 nm. Figure 18(e1)–18(e3) show the *x-z* ($y = 0$), *x-z* ($y = 35$ nm), and *x-z* ($y = 0$) plane images of $\sqrt{F_R(\lambda_{ex})}$ for $\phi$ of 0°, 90°, and 180°, respectively. All images exhibit the propagation of SE2PE light and the generation of SE2PE at the second HSs. Figure 18(f) shows the $\phi$ dependence of $\bar{F}_R(550\text{nm})$ at the second HSs, indicating that the SE2PE intensities fluctuate by approximately 15 times, depending on $\phi$. Thus, the $\phi$ dependence of $\bar{F}_R(550\text{nm})$ can be the origin of the variations in SE2PE intensities at the second HSs.

## IV. CONCLUSIONS

In this study, we investigated whether HSs between a silver NP and NW generates SE2PE, such as SHE-Ray, SEHRS, and two-photon SEF of dye molecules, with CW NIR laser excitation. We found the vibrational structures in $\sigma_{sca}(\lambda_{ex})$, large HS-by-HS variations in the SE2PE intensities, relationship of PER$_{SE2PE}$ > PER$_{SE1PE}$ in their polarization dependence, and SE2PE from neighboring HSs on a common NW. A comparison between these experiments and the FDTD calculations revealed that a large $F_R$ appears in the visible to NIR region owing to the EM coupling between the LPs of the NP and SPs of the NW. This ultrabroad resonance of the coupled plasmons enabled the detection of SE2PE light with CW laser excitation. The comparison also revealed that the vibrational structures in $\sigma_{sca}$ are the results of FP interference of the NWs transcribed via EM coupling, the large HS-by-HS variations are mainly induced by the $d_{gap}$ dependence of SE2PE intensities, the $PER_{SE2PE} > PER_{SE1PE}$ is the result of the complicated electric field distribution of quadrupole LP in the visible region involved in the EM coupling, and the SE2PE of the neighboring HS is induced by the propagation of SE2PE (not NIR excitation) light through the NW. We believe that this CW laser-excited nonlinear spectroscopy using EM enhancement of such HSs is applicable to various molecular spectroscopies related to plasmonics, such as the nonlinear counterparts of surface- and tip-enhanced spectroscopies.[41-43]

## ACKNOWLEDGMENTS

The authors thank Prof. Jeyadevan Balachandran (University of Shiga Prefecture) for providing the silver nanowires. This work was supported by a JSPS KAKENHI Grant-in-Aid for Scientific Research (C) (Grant No. 25K08520 and 25K08506).

## DATA AVAILABILITY

The data that support the findings of this study are available from the corresponding authors upon reasonable request.